\documentclass[fleqn,usenatbib]{mnras}

\usepackage{newtxtext,newtxmath}

\usepackage[T1]{fontenc}
\usepackage{ae,aecompl}

\usepackage{graphicx}	
\usepackage{amsmath}	
\usepackage{amssymb}	
\usepackage{array}

\newcommand{\ha}{H${\alpha}$}
\newcommand{\hb}{H${\beta}$}
\newcommand{\st}{[S\,{\sc II}]}
\newcommand{\nt}{[N\,{\sc II}]}
\newcommand{\zd}{[O\,{\sc III}]}
\newcommand{\zt}{[O\,{\sc II}]}

\title[An unusual transient following SGRB 071227]{An unusual transient following the short GRB 071227}

\author[R. A. J. Eyles et al.]{R. A. J. Eyles,$^{1}$\thanks{raje1@leicester.ac.uk}
P. T. O'Brien,$^{1}$
K. Wiersema,$^{2,1}$
R. L. C. Starling,$^{1}$
\newauthor
B. P. Gompertz,$^{2}$
G. P. Lamb,$^{1}$
J. D. Lyman,$^{2}$
A. J. Levan,$^{3,2}$
S. Rosswog$^{4}$
\newauthor
\& N. R. Tanvir$^{1}$ 
\\
$^{1}$Department of Physics and Astronomy, University of Leicester, University Road, Leicester, LE1 7RH, UK\\
$^{2}$Department of Physics, University of Warwick, Coventry, CV4 7AL, UK\\
$^{3}$ Department of Astrophysics/IMAPP, Radboud University, Nijmegen, The Netherlands\\
$^{4}$ The Oskar Klein Centre, Department of Astronomy, AlbaNova, Stockholm University, SE-106 91 Stockholm, Sweden
}

\date{Accepted XXX. Received YYY; in original form ZZZ}

\pubyear{2019}

\begin{document}
\label{firstpage}
\pagerange{\pageref{firstpage}--\pageref{lastpage}}
\maketitle

\begin{abstract}
We present X-ray and optical observations of the short duration gamma-ray burst GRB 071227 and its host at $z=0.381$, obtained using \textit{Swift}, Gemini South and the Very Large Telescope. We identify a short-lived and moderately bright optical transient, with flux significantly in excess of that expected from a simple extrapolation of the X-ray spectrum at 0.2-0.3 days after burst. We fit the SED with afterglow models allowing for high extinction and thermal emission models that approximate a kilonova to assess the excess' origins.
While some kilonova contribution is plausible, it is not favoured due to the low temperature and high luminosity required, implying superluminal expansion and a large ejecta mass of $\sim 0.1$ M$_{\sun}$. We find, instead, that the transient is broadly consistent with power-law spectra with additional dust extinction of $E(B-V)\sim0.4$ mag, although a possibly thermal excess remains in the \textit{z}-band.
We investigate the host, a spiral galaxy with an edge-on orientation, resolving its spectrum along its major axis to construct the galaxy rotation curve and analyse the star formation and chemical properties. The integrated host emission shows evidence for high extinction, consistent with the afterglow findings. The metallicity and extinction are consistent with previous studies of this host and indicate the galaxy is a typical, but dusty, late-type SGRB host.
\end{abstract}

\begin{keywords}
gamma-ray burst: general -- gamma-ray burst: individual: GRB 071227
\end{keywords}



\section{Introduction}

The detection of gravitational wave signal GW 170817 \citep{Abbott17}, an inspiral and merger of a binary neutron star system, coupled with the coincident detection of the short gamma-ray burst (SGRB) GRB 170817A by both the \textit{Fermi Gamma-ray Burst Monitor} (\textit{Fermi-GBM}) and \textit{INTErnational Gamma-Ray Astrophysics Laboratory} (\textit{INTEGRAL}) spacecraft \citep{Goldstein17,Savchenko17}, has greatly strengthened the link between SGRBs and a compact binary progenitor. The merger of two compact objects, either a binary neutron star (BNS) or a neutron star black hole (NSBH) system, is a cataclysmic event which can power a relativistic jet and an SGRB. This jet, and by extension, the gamma and X-ray emission, is emitted in the direction of the orbital axis of the progenitor system (`on-axis'). SGRBs are conventionally characterised by a $T_{90}$, the time in which 90\% of the prompt emission photons are detected, of less than 2~s and a hard gamma-ray spectrum \citep{Kouveliotou93}. There is a wide variety in SGRB host galaxies \citep{Leibler10,Fong13b}. This is in contrast to long gamma-ray bursts (LGRBs), which have a $T_{90}$ greater than 2~s and a softer gamma-ray spectrum, and are linked to highly energetic core collapse supernovae residing in blue star-forming, low metallicity galaxies \citep{Svensson10,Mannucci11}.

Both short and long GRBs produce broadband afterglows, which provide a great deal of information about the properties of the jet opening angles, energetics and circumburst medium \citep{Sari99a,Piran04}. Other types of transients are also associated with each type of GRB: for LGRBs, supernovae following the core collapse of their progenitor star \citep{Hjorth03}, and for SGRBs, kilonovae \citep[e.g.][]{Li98,Tanvir13,Metzger17}.

A kilonova is a radioactively powered and rapidly evolving transient, peaking in the optical or NIR depending on time, the progenitor properties and viewing angle \citep{Metzger17}. This is due to the behaviour of the different types of ejecta produced during and following the merger process: dynamical and disk wind ejecta \citep{Rosswog15}. The dynamical ejecta have a varying electron fraction ($Y_e$) with low $Y_e$ tidally ejected material distributed near the equatorial plane and high $Y_e$ collisional ejecta distributed in a symmetrical sphere at a smaller radius. The disk wind ejecta, similarly to the collisional ejecta, are also distributed in a symmetrical sphere around the merger site but, depending on the neutrino winds described later, have a low $Y_e$ and are found at even smaller radii.

The ejecta undergo {\em r}-process nucleosynthesis \citep{Lattimer74}, with the elements produced varying according to $Y_e$ \citep{Metzger17}. The neutron rich equatorial and disk wind material produces heavy lanthanides while the high $Y_e$ material is neutron poor, inhibiting lanthanide production.  Decay of the elements produced in this {\em r}-process nucleosynthesis leads to radioactive heating of the surrounding ejecta and the varying opacities of the ejecta results in the two key components of kilonova emission: $\sim$week long red infrared around the equator and more isotropically distributed $\sim$day long blue optical emission respectively. Following a typical SGRB, it would be expected that the on-axis viewing angle would mean that it is the blue component that would primarily be observed assuming it is not outshone by the SGRB afterglow itself, although it is also likely the red component would be observable as the viewing cone is broad. Depending on the masses and types of the progenitors, it is also possible for a magnetar central engine to form which produces a neutrino wind, increasing the $Y_e$ of the nearby disk wind ejecta. The additional energy injected by this magnetar, as well as the higher $Y_e$ of the disk wind ejecta, leads to a longer lived kilonova with bluer emission lasting several hours to days \citep{Kasen15}. In addition to these components, it has been proposed that kilonovae may be preceded by precursor emission peaking in the ultraviolet. Directly following the merger, the fastest ejecta can form an outer layer composed primarily of `free' neutrons. These free neutrons inject additional energy into the ejecta, heating the material and intensifying the emission produced \citep{Metzger15}.

Observational constraints on the diversity of kilonova behaviour are very limited \citep{Gompertz18}. For instance, it is believed that a kilonova was detected following GRB 130603B \citep{Tanvir13}, but this was limited to a single datapoint deviating from the underlying afterglow light curve. A number of other kilonova candidates are briefly discussed in comparison to GRB 071227 in section~\ref{sec:KN}.

The best example to date of a kilonova signature was the emission associated with the gravitational wave event GW 170817. The accompanying optical and infrared kilonova AT 2017gfo, has been observed in exquisite detail \citep[e.g.][]{Abbott17,Covino17,Cowperthwaite17,Kilpatrick17,Smartt17,Tanvir17}. This has allowed a much more thorough investigation into the resultant kilonova than has previously been possible. Both the red and blue emission can be seen which, along with the SGRB energetics and gravitational wave data, implies the kilonova is being viewed off-axis. The highly sampled light curves resulting from observations of AT 2017gfo also allow the temporal evolution of kilonova emission to be more accurately modelled \citep[e.g.][]{Gompertz18}.

There are only a few short bursts for which well-sampled multi-wavelength afterglows are available. These observables can be used to determine useful information on the properties of the progenitors and in turn allow the inference of global rates and properties of SGRBs. However, this is still a relatively small sample of SGRBs. GW 170817 has also prompted the evaluation of potential kilonovae in past SGRBs \citep{Gompertz18,Ascenzi19,Jin19,Rossi19,Troja19,Lamb19}. In this paper, we aim to bring the short GRB 071227 into both these catalogues.

GRB 071227 is a relatively well studied example of an SGRB. In particular, \citet{DAvanzo09} is a notable counterpart to this paper, presenting optical photometry of the GRB and photometric and spectroscopic analysis of the host as we do here. Their observations were performed using the FOcal Reducer and low dispersion Spectrograph (FORS2) instrument on the Very Large Telescope (VLT). They found GRB 071227 to be a fairly typical SGRB although they only had one observation during the afterglow phase. We revisit these observations in this paper and use them to obtain deeper upper limits than previously and in addition, we include new deep multi-band Gemini South observations. \citeauthor{DAvanzo09}'s spectroscopic observations of the host provided a redshift of 0.381, a star formation rate of 0.6 M$_{\sun}$~yr$^{-1}$ and a metallicity of $12 + \log(\text{O}/\text{H}) = 8.2 - 8.8$ dex . Additional optical and NIR photometry is found in \citet{DAvanzo07a}, \citet{Berger07a} and \citet{NicuesaGuelbenzu12}.

\citet{NicuesaGuelbenzu14} also analysed the host galaxy, presenting radio observations as well as identifying a \textit{Wide-field Infrared Survey Explorer} (\textit{WISE}) counterpart. Using spectral energy distribution (SED) models, they derive a star formation rate of $\sim24$ M$_{\sun}$\,yr$^{-1}$. The contrast with \citet{DAvanzo09}'s results is suggested to be due to high optical extinction in the host. Further SED analysis is also presented by \citet{Leibler10} who measure stellar population ages. The host morphology is investigated in \citet{Fong13b}.

GRB 071227 has also been evaluated as a GRB with extended emission (EE) \citep{Gompertz14,Lien16,Gibson17}, finding it to fit well with a magnetar propeller model, while \citet{Kisaka17} found evidence of both extended and plateau emission, and has been investigated in relation to X-ray flares in SGRBs \citep{Margutti11}, in which a flare was identified in the first 200 seconds following the GRB. Finally, GRB 071227 is in the sample examined by \citet{Rossi19} for kilonova signatures by comparison with AT 2017gfo, finding that any kilonova is likely several times dimmer than the afterglow.

In this paper, we report our own optical photometry obtained using the Gemini Multi-Object Spectrograph instrument (GMOS-S) at the Gemini South observatory, finding, at a time of $\sim$0.26 days, apparent excess optical flux when compared with a synchrotron afterglow model extrapolation of the X-ray emission. We investigate the scenarios that could lead to this apparent excess, including kilonovae and reverse shocks. We also add further spectroscopic observations and analysis thereof of the host. In section 2, we summarise our Gemini South observations and processing as well as the available data from the \textit{Neil Gehrels Swift Observatory}, hereafter \textit{Swift}, satellite and the Very Large Telescope. Section 3 deals with the properties of the X-ray and optical transients and our investigation into the mechanisms that could have produced them, while section 4 examines the properties of the host galaxy. Section 5 presents our conclusions.

Throughout this paper we adopt a cosmology with $H_0 =  71$ km\,s$^{-1}$\,Mpc$^{-1}$, $\Omega_m = 0.27$ and $\Omega_\Lambda = 0.73$. A redshift of $z = 0.381$ therefore corresponds to a luminosity distance of 2039.0 Mpc, and 1 arcsecond corresponds to 5.4 kpc. Hereafter we use the notation $F \propto t^{-\alpha} \nu^{-\beta}$, where $\alpha$ and $\beta$ are the temporal and spectral power law indices, respectively. In the case of spectral fits performed using \textsc{xspec}, our errors are 90\% confidence. All other errors are given to 1 $\sigma$.

\section{Observations and Data Reduction}

\begin{figure*}
\includegraphics[width=85mm]{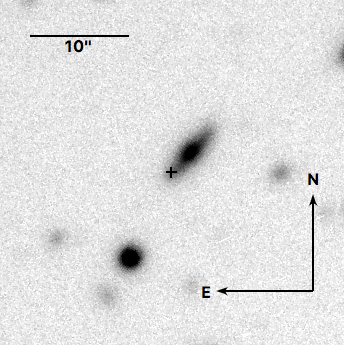}
\includegraphics[width=85mm]{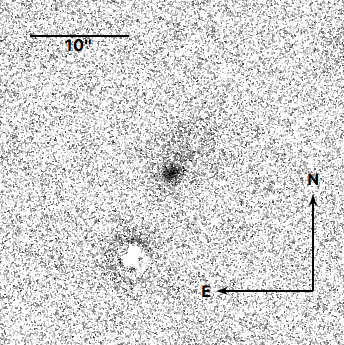}
\caption{The observed field near GRB071227 in epoch 1 in the \textit{r} filter (left) and the same field with epoch 4 subtracted (right), demonstrating the fading optical transient. The cross in the left image indicates the position of the transient in the host. A subtraction artifact (the residual from a bright star) is also visible below the residual from the transient.}
\label{fig:SubtractionEx}
\end{figure*}

\subsection{\textit{Swift} observations \label{sec:swiftobs}}

\textit{Swift} was triggered at 20:13:47 UT and slewed to the burst at RA 03h52m29s, Dec -55d57m08s \citep{Sakamoto07a}. The prompt light curve indicated multiple peaks during a $T_{90}$ of $1.8 \pm 0.4$s \citep{Sakamoto07b}. However, a more careful reduction revealed a lengthy extended emission period following the burst, in common with some other SGRBs. If accounted for, the extended emission increases the $T_{90}$ estimate considerably to 142.5 $\pm$ 48.4 s \citep{Lien16}. In the \textit{Swift}-BAT bandpass, $E_{\gamma,\text{iso}}$ is $1.96^{+0.30}_{-0.29}\times 10^{50}$ erg.

We retrieved the XRT data from the UK \textit{Swift} Science Data Centre (UKSSDC)\footnote{\url{http://www.swift.ac.uk/index.php}} and corrected it for line-of-sight absorption using the ratio of counts-to-flux unabsorbed to counts-to-flux observed from the fit to the late-time photon counting mode spectrum on the UKSSDC. This is a factor of $1.02$.

\subsection{GMOS-S imaging \label{section_GSobs}}

\begin{table*}
 \centering
  \caption{Log of Gemini South observations. All magnitudes are AB and corrected for Galactic extinction. The upper limits are 3 $\sigma$.}
  \label{tab:ObsLog}
  \begin{tabular}{@{}ccccccccc@{}}
  \hline
   Epoch & Time started & Mean time after  & Exposure & Filter/grism & Seeing (mean & Mean airmass &  OT magnitude \\ 
    & (UT) &  burst (days) & time (s) & & FWHM in arcsec) & & & \\
 \hline
  1 & Dec 28 2007 00:56:58 & 0.27119 & 20$\times$180 & \textit{r} & 1.75 & 1.16 & 24.03 $\pm$ 0.04\\
    &                 & 0.26675 & 10$\times$180 & \textit{i} & 1.56 & 1.11 & 23.26 $\pm$ 0.04\\
    &                 & 0.23900 & 10$\times$180 & \textit{z} & 1.54 & 1.11 & 22.47 $\pm$ 0.07\\  
  2 & Jan 01  2008 02:06:37 & 4.30379 & 20$\times$180 & \textit{r} & 0.86 & 1.19 & $\gid 26.72$
  \\
    &                 & 4.32089 & 10$\times$180 & \textit{i} & 0.77 & 1.20 & $\gid 25.23$
    \\
    &                 & 4.29318 & 10$\times$180 & \textit{z} & 0.71 & 1.15 & $\gid 24.10$ 
    \\
  3 & Jan 04  2008 05:08:14 & 7.39899 & 20$\times$180 & \textit{r} & 0.97 & 1.54 & $\gid 26.85$ 
  \\
    &                 & 7.43925	 & 3$\times$180  & \textit{i} & 0.92 & 1.78 & --- \\
    &                 & 7.43092 & 3$\times$180  & \textit{z} & 1.48 & 1.86 & ---\\
  4 & Jan 11 2008 02:40:48 & 14.29787 & 20$\times$180 & \textit{r} & 0.98 & 1.21 & --- \\
\hline
  5 &  Jan 17 2008 00:56:21 & 20.19667 & 1830 & GG455\_G0329 & 0.77 & 1.11 & Spectrum \\ & & & & / R400  &  &  &   \\  
\hline
\end{tabular}
\end{table*}

Optical observations were performed at Gemini South using the Gemini Multi-Object Spectrograph (GMOS-S) \citep{Hook04}. These observations were performed using the original EEV CCDs, with an unbinned pixel size of 0.073 arsec. In this case, 2x2 binning was used, hence our pixel size is 0.146 arcsec. The \textit{r}, \textit{i} and \textit{z} filters used in GMOS-S are based on those used in the Sloan Digital Sky Survey as presented in \citet{Fukugita96}. Conditions during the observations were generally good, although seeing during the first epoch was notably poorer (see Table \ref{tab:ObsLog}).

We reduced the data using the \textsc{gemini iraf} package, specifically the \textsc{gmos} subpackage. The method used is based on that found in the GMOS Data Reduction Cookbook \citep{Shaw16}. Master bias and flat frames were constructed for each filter using the available observations closest to each epoch and were then applied to the raw science frames. There was significant fringing present in the raw \textit{i} and \textit{z} frames, a common problem for images taken using the older EEV CCDs. Typically, the fringing is removed using master fringe frames taken once per semester. However, this was found to be ineffective in this case and instead we constructed fringe frames from the science frames using the \textsc{gifringe} task. These were found to be much more effective at minimising the fringing present in the \textit{i} and \textit{z}. As there was no apparent fringing in the \textit{r} frames, no correction was applied. The data from the individual CCDs were then mosaiced together to create an image of the entire field. The median value of the data was substituted into any blank pixels, such as those in the gaps between the CCDs.

We combined the individual images using {\sc IRAF} tasks {\sc shiftadd}\footnote{Developed by E. Rol.} and {\sc imcoadd}, with the final output being the mean of the stack. We used the \textsc{gaia} software \citep{Draper14} to produce a final world coordinate system (WCS) calibration for each image manually using 49 USNO-B objects in the field and derived a photometric calibration for each stacked frame by matching stars to the SkyMapper catalogues \citep{Wolf18}.

We used the \textsc{hotpants} code \citep{Becker15} for image subtraction, a reimplementation of the \textsc{isis} algorithm \citep{Alard00}. This uses a space-varying kernel method to achieve effective subtraction across the entire field by matching point spread functions (PSFs) between images. The last epoch observed for each filter were taken as the template images, i.e. epoch 4 for \textit{r} and epoch 3 for \textit{i} and \textit{z}. The subtraction for epoch 1 in the \textit{r}-band is shown in Figure~\ref{fig:SubtractionEx}.

\subsection{VLT/FORS2 imaging}

\begin{table*}
 \centering
  \caption{Log of VLT/FORS2 observations. All observations used the ESO \textit{R\_Special} filter. All magnitudes are AB and corrected for Galactic extinction. The upper limits are 3 $\sigma$.}
  \label{tab:VLTLog}
  \begin{tabular}{@{}cccccccc@{}}
  \hline
   Epoch & Time started & Mean time after & Exposure & Mean seeing & Mean airmass &  OT magnitude \\ 
    & (UT) &  burst (days) & time (s) & (arcsec) & & & \\
 \hline
  1 & Dec 28 2007 03:08:37 & 0.29055 & 2$\times$120  & 0.7 & 1.20 & 24.17 $\pm$ 0.12\\
  2 & Dec 31 2007 05:05:38 & 3.37133 & 3$\times$180  & 0.7 & 1.46 & $\gid 25.39$ \\
  3 & Jan 03 2008 02:30:56 & 6.26790 & 5$\times$180  & 0.8 & 1.19 & $\gid 24.83$ \\  
  4 & Jan 07 2008 02:02:10 & 10.24790 & 5$\times$180  & 0.8 & 1.18 & $\gid 25.01$ \\
  5 & Jan 16 2008 05:16:31 & 19.38287 & 5$\times$180  & 0.7 & 1.84 & $\gid 25.23$\\
  6 & Jan 18 2008 01:58:14 & 21.25139 & 10$\times$180  & 0.8 & 1.22 & $\gid 25.32$ \\
  7 & Jan 23 2008 02:12:35 & 26.21775 & 10$\times$180  & 0.9 & 1.19 & --- \\
  8 & Feb 06 2008 02:12:35 & 40.26121 & 10$\times$180   & 1.1 & 1.41 & --- \\
\hline
\end{tabular}
\end{table*}

As previously mentioned, we have reanalysed the VLT data examined in \citet{DAvanzo09}. Observations were performed using the \textit{R\_Special} filter \citep{ESO2018} on the FORS2 instrument, which has an unbinned pixel size of 0.125 arcsec. Again, 2x2 binning was used leading to a pixel size of 0.25 arcsec. The data were obtained from the European Southern Observatory archive\footnote{\url{http://archive.eso.org/cms.html}}.

We processed the data using the standard \textsc{EsoReflex} workflow for the FORS instruments \citep{Freudling13} then stacked the individual frames using \textsc{imcoadd} as described above. Photometric calibrations were again evaluated by comparison with the SkyMapper catalogue \citep{Wolf18}, calibrating directly to the \textit{r}-band. Image subtraction was performed using \textsc{hotpants}, again taking the final epoch, in this case epoch 8, as the template image. It should be noted that we were unable to achieve an effective subtraction for epoch 7, likely due to this image being of lower quality and \textsc{hotpants} therefore being unable to accurately match the PSFs.

\subsection{GMOS-S spectroscopy \label{section_GSobs_spec}}

\begin{figure}
\centerline{\includegraphics[width=8.5cm]{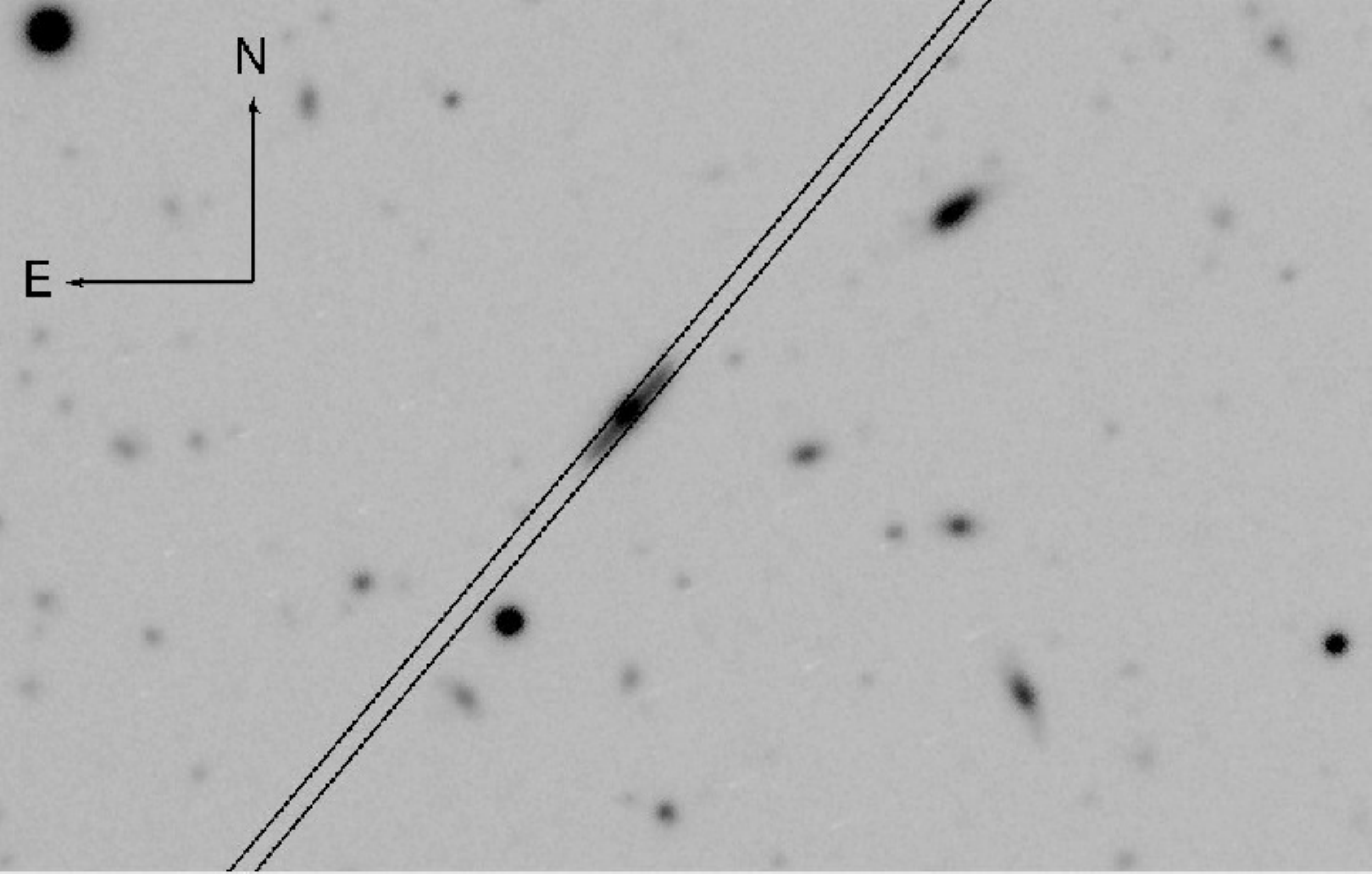}}
\centerline{\includegraphics[width=8.5cm]{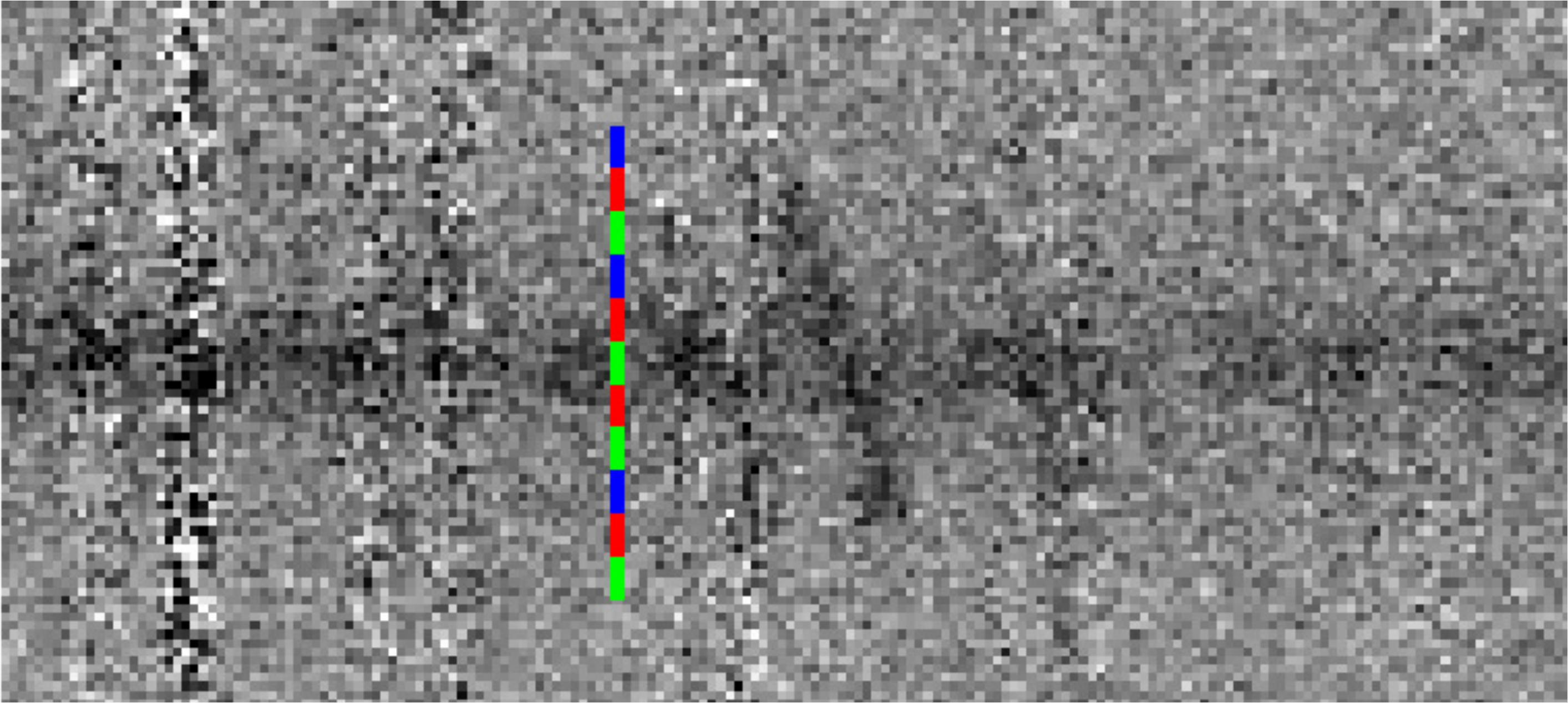}}
\caption{The orientation of the 1.0 arcsecond wide GMOS-S slit, 140 degrees parallactic angle, is indicated on the second epoch $r$ image (top) and a small part of the 2 dimensional spectrum  centered on the \ha\ emission line(bottom). Redwards (right) of the \ha\ line the, much fainter, \nt\ $\lambda$6585 line is visible. The total aperture consisting of 11 sub-apertures of each 5 pixels is indicated. The noisy vertical stripes perpendicular to the trace of the galaxy are low amplitude residuals from the skyline subtraction through the nod \& shuffle technique.}
\label{fig:apperture}
\end{figure}

To further characterize the host galaxy, we obtained a spectrum with GMOS-S using the nod and shuffle technique, which increases signal to noise at the red end by increasing accuracy of skyline subtraction (see \citet{Glazebrook01} for a description of the technique) and suppresses the effects of the severe fringing of the GMOS-S detectors. The 1.0 arcsecond wide slit was oriented along the major axis of the galaxy (sky position angle 140 degrees, see Figure~\ref{fig:apperture}), in contrast to \citet{DAvanzo09} who oriented their slit North to South centred on the nucleus of the host. We used low resolution grism R400 and the GG455\_G0329 filter, resulting in a wavelength range of $\sim$5500 to 9290 \AA\ (with a central wavelength of $\sim$7200 \AA). 30 nod and shuffle cycles were used and the total net exposure time was 1830 seconds.  An atmospheric dispersion compensator was used to minimize colour dependent slit-losses.  The data were reduced using the standard nod \& shuffle procedures in the {\sc gemini} package in {\sc IRAF}, and tasks in the {\sc specred} package were used for extraction.

We extracted the host galaxy spectrum by using the relatively bright continuum of the bulge to fit the shape of the trace, then extracted 11 adjoining, equally sized subapertures following this trace. The subapertures are 5 pixels in size, corresponding to 0.73 arcseconds per subaperture, i.e. a value broadly matched to the seeing FWHM. This corresponds to a physical scale of 3.9 kpc per subaperture. The GRB location falls in subaperture 2. The spectra extracted from each subaperture are referred to as subspectra hereafter. The subspectra were wavelength calibrated using a CuAr lamp spectrum and the dispersion solution had an RMS of 0.3 \AA. From the full-width-at-half-maximum (FWHM) of a Gaussian fit on the arc lines, we measured a nominal spectral resolution of 6.9 \AA, corresponding to 280 km/s at 7300 \AA.    

Flux calibration of the subspectra was done using observations of the spectrophotometric standard star Hiltner 600 (\citealt{Hamuy92}; \citealt{Hamuy94}), taken under photometric conditions. Atmospheric extinction correction was done by applying the average CTIO atmospheric extinction curve. As the effective airmass was low, the effects of the atmospheric extinction correction were negligible. A Galactic dust extinction correction was performed by using the  $E(B-V)$ value of 0.013 mag \citep{Schlegel98}, assuming a Galactic extinction law $A_{\lambda}/A_{V}$ expressed as $R_{V} = A_{V}/E(B-V)$ \citep{Cardelli89}. We made the standard assumption $R_{V} = 3.1$ \citep{Rieke85}.

\section{Properties of the transient}

 \subsection{Broadband light curve}
\label{sec:broad_lc}

\begin{figure*}
\includegraphics[width=17cm]{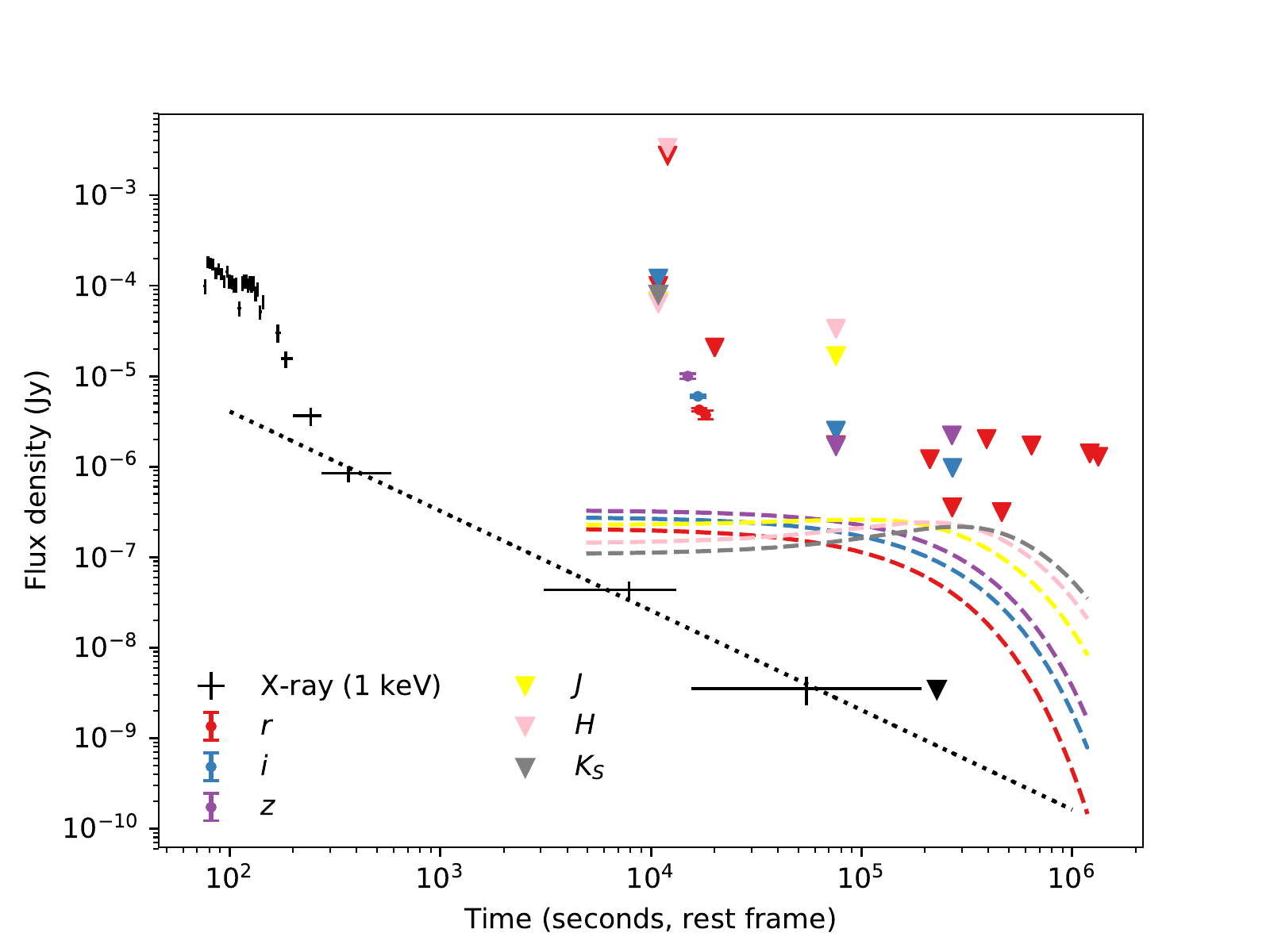}
\caption{The broadband light curve of GRB 071227. Extinction corrections of $N_H=2.97\times 10^{21}$ cm$^{-2}$ and $E(B-V)=0.43$ mag have been applied. Our \textit{Swift} and optical data are supplemented by additional optical and NIR photometry from GROND \citep{NicuesaGuelbenzu12}, REM \citep{DAvanzo07a} and Magellan \citep{Berger07a}. Triangles represent 3 $\sigma$ upper limits and the black dotted line indicates the slope of $\alpha = 1.1$ followed in the X-ray. Models of the kilonova AT 2017gfo light curves \citep{Gompertz18}, extrapolated to the redshift of GRB 071227, are shown in the dashed lines.
}
\label{fig:lcplot}
\end{figure*}

\begin{figure}
\includegraphics[width=8.5cm]{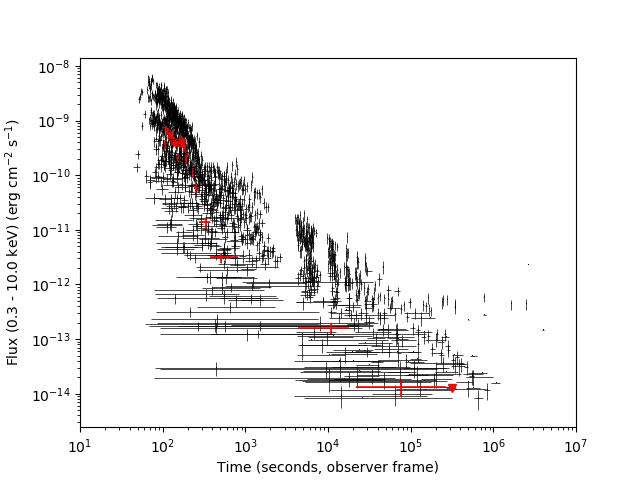}
\caption{The available XRT light curves for the SGRBs in the catalogue presented in \citet{Fong15} in black with GRB 071227 highlighted (red).}
\label{fig:XRTlc_comparison}
\end{figure}

The \textit{Swift} X-Ray Telescope (XRT) identified an X-ray transient at RA 03h52m31.21s, Dec -55d59m03.1s \citep{Beardmore07}. The XRT light curve, as shown in black in Figure~\ref{fig:lcplot} and red in Figure~\ref{fig:XRTlc_comparison}, is well fitted with a double broken power law between 210 s and 34.8 ks. This fit has parameters $\alpha_1=1.1 \pm 0.2$, $t_{\text{break}_1}=183 \pm 7$s, $\alpha_2=5.3^{+1.2}_{-0.6}$, $t_{\text{break}_2}=389 \pm 65$s and $\alpha_3=1.1 \pm 0.2$ \citep{Beardmore07,Evans07,Evans09}. When compared to most \textit{Swift} detected SGRBs, as shown in Figure~\ref{fig:XRTlc_comparison}, the XRT light curve is found to be slightly under luminous, particularly at later times, but is not atypical. 

We used SExtractor 2.19.5 \citep{SExtractor96} to detect and extract sources from our optical data to a 5 $\sigma$ confidence threshold. Using the first epoch \textit{r}-band difference image, we identified an optical transient which had an AB magnitude of 24.03 $\pm$ 0.03 at a position of RA 03h52m31.25s, Dec -55h59m02.87s (J2000) with an uncertainty of 0.28 arcseconds dominated by the RMS of the astrometric solution of the USNO stars used for the WCS calibration. This is consistent with the positions of the X-ray and optical transients found by \citet{Beardmore07} and \citet{DAvanzo09} respectively. The transient is also detected in the first epoch \textit{i} and \textit{z}-band images at AB magnitudes 23.26 $\pm$ 0.04 and 22.47 $\pm$ 0.07, respectively.

The optical transient is not detected in subsequent epochs, however, indicating its fading. We identified upper limits for these later epochs by injecting artificial sources into the input image and performing subtraction to identify where the magnitude error reached the 3 $\sigma$ threshold. The detected magnitudes and upper limits from our GMOS-S observations are summarised in Table~\ref{tab:ObsLog}.

This rapid fading can also be identified in our analysis of the VLT/FORS2 data as summarised in Table~\ref{tab:VLTLog}. Again, the transient was only identified in the first epoch, at an \textit{R}-band AB magnitude of 24.17 $\pm$ 0.12, and we therefore again used injected artificial sources in the later epochs to evaluate deeper upper limits than those identified by \citet{DAvanzo09}. As we were unable to obtain an effective subtraction, we do not have an upper limit for epoch 7. Fitting our detections and upper limits with power laws indicated minimum temporal decay indices of $\alpha_r \ga 0.87$, $\alpha_i \ga 0.65$ and $\alpha_z \ga 0.52$ for the \textit{r}, \textit{i} and \textit{z} bands respectively.

\subsection{Broadband spectral energy distribution (SED)}
\label{sec:SEDs}

We used \textsc{xspec 12.10.0} \citep{Arnaud96,Dorman01} from \textsc{HEASoft 6.24} to evaluate the broadband SED. Full details on the models used and our fitted parameters are presented in Appendix \ref{app:SED_detail}. We first fitted the X-ray and optical data separately. An absorbed power law model was used to fit the XRT spectrum at a mean time in log space coincident to the first optical GMOS-S epoch, 21635s after the burst, finding a photon index, $\Gamma=\beta+1$, of $1.58^{+0.87}_{-0.57}$ and host intrinsic column density upper limit of $N_H \lid 2.97 \times 10^{21}$ cm$^{-2}$ in addition to the fixed Galactic column density of $N_H = 1.31 \times 10^{20}$ cm$^{-2}$ \citep{XSPECgalcolumn} (see Table \ref{tab:Xray_spec} for full details). This fit has a reduced chi-squared of 1.130. The parameters identified agree with those found by \textit{Swift}'s automatic fitting algorithms \citep{Evans09} and are consistent with the broader sample of SGRBs \citep{DAvanzo14,Fong15}.

Both the light curve and spectral properties of the X-ray data are consistent with a typical afterglow model synchrotron emission from a relativistic blast wave \citep{Sari99a,Sari99b}. The photon index and the temporal slope ($\alpha_3$ at this time) can be used to independently evaluate the electron energy distribution index, $p$, using the standard closure relations \citep{Sari98,Chevalier99,Chevalier00,Granot02}:
\begin{equation}
p(\Gamma)=\begin{cases}
2\Gamma -1 = 2.16^{+1.74}_{-1.14}  & \text{for $\nu_m<\nu<\nu_c$}\\
2(\Gamma -1) = 1.16^{+1.74}_{-1.14}  & \text{for $\nu_m<\nu_c<\nu$}
\end{cases}
\label{eq:p_gamma}
\end{equation}

\begin{equation}
p(\alpha,k)=\begin{cases}
\frac{4\alpha + 3}{3} = 2.47 \pm 0.27  & \text{for $\nu_m<\nu<\nu_c$ and $k=0$} \\
\frac{4\alpha + 1}{3}  = 1.80 \pm 0.27 & \text{for $\nu_m<\nu<\nu_c$ and $k=2$} \\
\frac{2(2\alpha + 1)}{3} = 2.13 \pm 0.27 & \text{for $\nu_m<\nu_c<\nu$ and any $k$} 
\end{cases}
\label{eq:p_alphanok}
\end{equation}
where $\nu_m$ is the peak frequency, $\nu_c$ is the cooling frequency and $k$ describes the particle number density of the surrounding medium as a function of radius, $n(r)\propto r^{-k}$, although it should be noted that a wind like density profile is not expected following a neutron star merger, hence $k$ is expected to be 0 \citep{Chevalier99}. We do not include energy injection which is thought to be responsible for the unusual features observed in a subset of SGRB light curves \citep[e.g.][]{Gompertz14}. Together, these imply $2.1 \la p \la 2.4$ in the regime $\nu_m<\nu<\nu_c$, consistent with the results of \citet{Fong15} where they found $p = 1.92\pm 0.31$ for this burst and a mean value of $p$ of  $2.43^{+0.36}_{-0.28}$ for their full population of SGRBs.

\begin{figure}
\includegraphics[width=8.5cm]{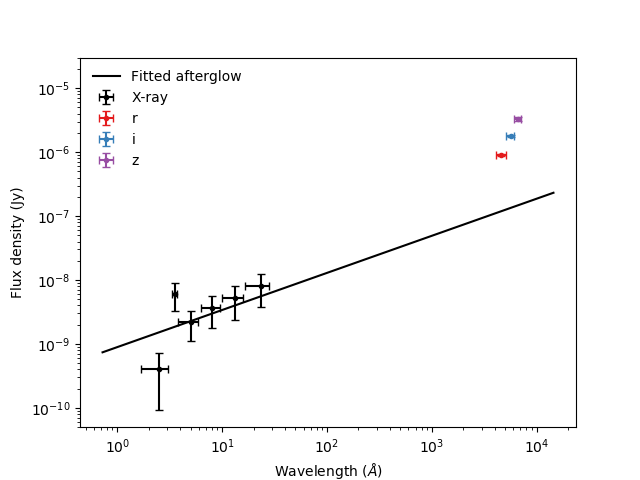}
\caption{The X-ray only afterglow fit extrapolated to optical wavelengths demonstrating the apparent optical excess.}
\label{fig:xrayonlysed}
\end{figure}

\textsc{Xspec} was also used to fit our first GMOS-S optical epoch. Because the mean time of our \textit{z}-band observation was somewhat earlier than the other bands, we extrapolated the flux density to this time assuming a power law decay with temporal decay index of 1.1, consistent with that inferred from the X-ray light curve. We again used an absorbed power law model to find $\Gamma\sim 4.49$ with no host extinction contribution assumed (see Table \ref{tab:Optical_spec} for full details). While this is somewhat unconstrained due to only having three datapoints available, this is still a surprising result, as the optical spectrum is typically shallower than the X-ray at this time. It also implies a much higher electron energy distribution index than typically found in GRBs ($\sim7 - 8$ depending on the regime assumed). Extrapolating the power law derived from the X-ray SED to optical wavelengths, as shown in Figure~\ref{fig:xrayonlysed}, also indicates that the optical flux is underestimated, even when the effects of optical extinction have not been removed. Below, we consider several possibilities for the apparent discrepancy.

\subsubsection{An afterglow explanation for the apparent optical excess}

It is possible that the optical and X-ray transients are indeed consistent with a single afterglow model. We therefore attempted to fit both the X-ray and optical data together using \textsc{xspec} to gain a better picture of the varying situation between these wavelengths. Again, an absorbed power law model was used, this time a combination of both the X-ray and optical models. A pair of power law fits with a reduced chi-squared of 1.113 and the same photon index of $1.58^{+0.56}_{-0.57}$ were found, which required both a high optical extinction ($E(B-V) \sim 0.85$) and the normalisation of the optical to be $\sim$140 times larger than that of the X-ray (see Table \ref{tab:Combined_spec} for full details), with any single normalisation fit being extremely poor. This level of extinction is plausible in the host, as we find in section~\ref{sec:Spectroscopy}, but the large difference in the normalisation is still problematic.

\begin{figure}
\includegraphics[width=8.5cm]{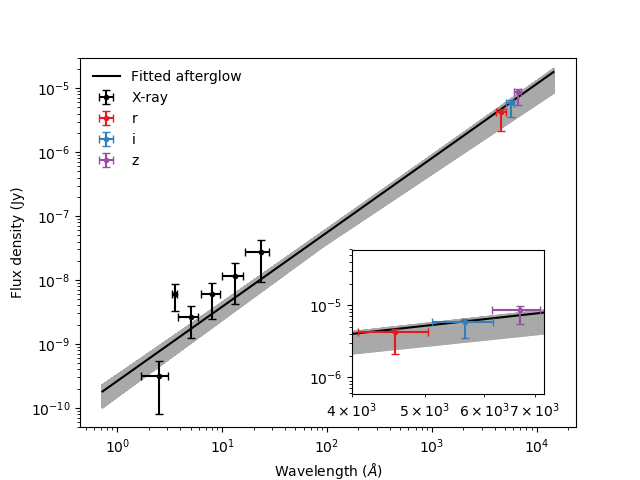}
\includegraphics[width=8.5cm]{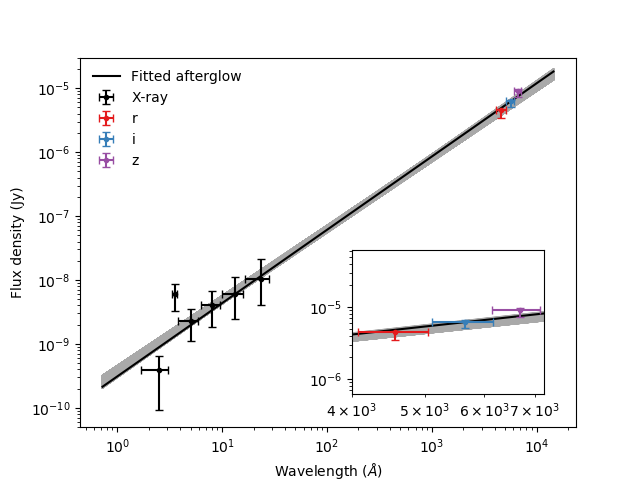}
\caption{Our afterglow fits obtained using an MCMC method where $N_H$ and $A_V$ are linked by the relation identified by \citet{Guver09} (top) and where they are free to vary independently (bottom). The shaded regions indicate the 1 $\sigma$ errors on our fits.
}
\label{fig:MCMC_fits}
\end{figure}

Due to limitations within \textsc{xspec} (the large wavelength difference between our X-ray and optical data and the greater amount of X-ray data meaning any fit is naturally weighted towards fitting best with the X-ray), we used a Markov Chain Monte Carlo (MCMC) method to better evaluate the afterglow. The X-ray data were rebinned and the errors of our X-ray only fit were used to define our fit parameter space. We first assumed that the $N_H$ and $A_V$ were related by $N_H = (2.21 \pm 0.09) \times 10^{21} A_V$, as identified by \citet{Guver09} for the Galaxy. Our fitted parameters, therefore, were the slope and normalisation of the afterglow and the $N_H$ with $E(B-V)$ derived from the relation above. Assuming $R_V=3.1$ as in the Galaxy, this identified a fit to a $\Gamma$ of $2.19^{+0.06}_{-0.04}$ at a reduced chi-squared of 2.501, where $N_H = 2.97^{+0.01}_{-0.64} \times 10^{21}$ cm$^{-2}$ and $E(B-V) = 0.43^{+0.01}_{-0.09}$ mag (see Table \ref{tab:MCMC_tied} for full details), as shown in Figure \ref{fig:MCMC_fits} (top). Using other values of $R_V$, such as 2.93 and 3.16 as found in the Small and Large Magellanic Clouds (SMC and LMC) respectively, has an insignificant effect on this fit. The extinction corrections identified in this fit have been applied in Figure \ref{fig:lcplot}.

In a number of GRBs, the relation between $N_H$ and $A_V$ above is not obeyed, however, likely due to a dust to gas ratio in the host that differs from that of the Milky Way \citep[e.g.][]{Li08a,Li08b}. We therefore also fitted our SED allowing $E(B-V)$ to vary independently of $N_H$, as shown in Figure \ref{fig:MCMC_fits} (bottom). This identified a fit to a $\Gamma$ of $2.15^{+0.01}_{-0.06}$, where $N_H = 2.77^{+5.51}_{-2.77} \times 10^{20}$ cm$^{-2}$ and $E(B-V) = 0.45^{+0.01}_{-0.07}$ (see Table \ref{tab:MCMC_untied} for full details). This fit has a reduced chi-squared of 3.094, an increase likely due to the extra parameter being fitted. This result possibly implies a high dust to gas ratio in the host relative to the Milky Way, but could also indicate that the fit is relatively insensitive to $N_H$.

The above fits both still result in a \textit{z}-band excess, but are generally consistent within errors and are also consistent with the results from \textit{Swift}'s automatic fitting. Returning to the electron energy distribution index, we find that for the first of our MCMC fits, where $N_H$ and $A_V$ are related as in the Galaxy, that:
\begin{equation}
p(\Gamma)=\begin{cases}
2\Gamma -1 = 3.38^{+0.12}_{-0.08}  & \text{for $\nu_m<\nu<\nu_c$}\\
2(\Gamma -1) = 2.38^{+0.12}_{-0.08}  & \text{for $\nu_m<\nu_c<\nu$}
\end{cases}
\label{eq:p_gamma_refitted}
\end{equation}
This, together with our analysis of $p$ in relation to $\alpha$ above, still indicates that $2.1\la p \la 2.4$, but in the regime $\nu_m<\nu_c<\nu$. This result is also still consistent with the findings of \citet{Fong15}.

\subsubsection{A kilonova explanation for the apparent optical excess}
\label{sec:KN}

An alternate and possibly the more obvious source for apparent excess flux at optical or IR wavelengths following a short GRB is thermal emission arising from a kilonova. However, we find this to be somewhat problematic in the case of GRB 071227. Taking our X-ray only fit ($\Gamma\sim 1.58$) to be representative of the afterglow, fitting the remaining excess with no reddening assumed to a thermal blackbody model yields a low temperature ($\sim3600$ K) but a high luminosity ($\sim 1.3\times10^{43}$ erg s$^{-1}$) that require a superluminal expansion velocity according to the Stefan-Boltzmann law. 

The ejecta mass required for such a luminosity is also somewhat problematic in itself. Simply scaling the luminosity inferred from AT 2017gfo implies an ejecta mass following GRB 071227 of $\sim0.2 $ M$_{\sun}$, double the approximate maximum expected for a BNS merger of $\sim0.1 $ M$_{\sun}$ \citep{Metzger17}. 

Alternatively, we can directly estimate the ejecta mass by inspecting nuclear heating rates. Using the Finite Range Droplet Model \citep[FRDM;][]{Moller95} for the nuclear masses, network calculations indicate a nuclear heating rate of $\sim10^{11}$ erg g$^{-1}$ s$^{-1}$ at 0.26 days \citep{Korobkin12}. Making the extreme assumption that all released energy is observed as radiation yields a lower limit on the ejecta mass of 0.09 M$_{\sun}$. By applying a more realistic efficiency of $\sim 0.6$ \citep[e.g. Fig 8 in][]{Rosswog17}, however, indicates a mass closer to the $\sim0.2 $ M$_{\sun}$ found above. It is also important to realize that the nucleosynthesis path for the very low electron fractions of a neutron star merger meanders through largely unknown nuclear territory close to the neutron-dripline where no experimental information is available. Theoretical nuclear mass formulae that equally well reproduce known nuclear data therefore, can yield substantially different heating rates. For example, very low $Y_e$ trajectories produce substantially different amounts of nuclei in the trans-lead region ($A > 200$) depending on whether, for instance, the FRDM mass model or the Duflo-Zuker mass formula \citep[DZ31;][]{Duflo95} are employed. As a result, the effective heating rates at time scales of days can differ by factors of a few \citep{Barnes16,Rosswog17}. Making the most optimistic assumptions, namely that the ejecta are dominated by very low $Y_e$ material and the nuclear heating is close to the  DZ31 mass formula values, the total ejecta mass may be brought down below $\sim 0.1$ M$_{\sun}$. Such large values can in principle be ejected in some neutron star black hole mergers for favorable parameter combinations (mass ratio, neutron star compactness, black hole spin) \citep{Rosswog05,Foucart14,Brege18}, but not all neutron star black hole mergers are expected to eject such large masses \citep[e.g.][]{Foucart19}. In summary, the large required mass may be ejected in a neutron star black hole merger, but it may require a collusion of number of parameters and therefore, while in principle possible, is not our preferred interpretation.

There are two other mechanisms that could reduce the ejecta mass required, a `free' neutron precursor \citep{Metzger17} or a trapped jet leading to a bright thermally emitting relativistic cocoon peaking at optical wavelengths \citep{Kasliwal17}. However, a UV precursor is inconsistent with the much redder temperatures observed here, while the gamma-ray and X-ray data both show no indication of such a trapped jet, in which case any jet heating is too inefficient to produce the high luminosity seen here \citep{Duffell18}.

\begin{figure}
\includegraphics[width=8.5cm]{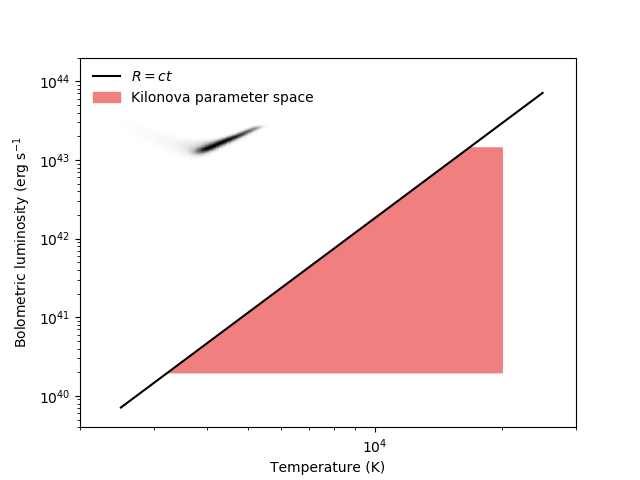}
\caption{The approximate parameter space open to kilonovae (light red) where the maximum luminosity at lower temperatures is defined by the maximum radius to which the ejecta velocity can expand by the time of our first optical epoch (black line). The greyscale indicates the range of our blackbody fits as the afterglow parameters and optical extinction are varied.}
\label{fig:KN_paramspace}
\end{figure}

Although this fit is unsuccessful, the errors on our afterglow and optical extinction expand the available parameter space. Varying the afterglow and extinction and refitting the excess, therefore, provides a more thorough investigative tool into the possibility of this excess being the result of a kilonova. For a given temperature, the luminosity of a kilonova is limited such that the expansion velocity remains below $c$. This is given by:
\begin{equation}
L < {4 \pi c^2 t^2 \sigma T^4}
\end{equation}
where $t$ is the time since the merger in the rest frame and $\sigma$ is the Stefan-Boltzmann constant. Along with assuming a maximum ejecta mass of 0.1 $M_{\sun}$, an \textit{r}-process heating rate of $\sim10^{11}$ erg s$^{-1}$ g$^{-1}$ as derived using the FRDM and an observed temperature range of 3000 to 20000 K, this defines our expected kilonova parameter range, shown in Figure \ref{fig:KN_paramspace}. Also shown in Figure \ref{fig:KN_paramspace} in greyscale are the parameters inferred by fitting blackbodies to the optical excess as the extinction, normalisation and slope of the afterglow power law are varied within the full range of the 90\% errors inferred by our X-ray only fit. We again used the relation identified by \citet{Guver09} to infer an upper limit to the optical extinction of $E(B-V) \sim 0.45$ from the fitted $N_H$ upper limit.

The continued lack of agreement between these fits and the plausible parameter space of our kilonova is clear and strongly indicates that the optical excess following GRB 071227 is not primarily the result of a kilonova and is instead likely dominated by afterglow. A thermal contribution is still plausible, however, which could be responsible for the small excess still apparent in the \textit{z}-band of our afterglow fits and it is therefore useful to compare the properties of GRB 071227 with several kilonova candidates: GRBs 050709 \citep{Hjorth05b,Fox05,Jin16}, 080503 \citep{Perley09,Metzger14a,Metzger14b}, 130603B \citep{Tanvir13} and 150101B \citep{Fong16,Troja18b}. Other candidates include GRBs 070809 \citep{Jin19} and 160821B \citep{Troja19,Lamb19}, although these are more recently identified and therefore not examined in detail here.

\begin{figure}
\includegraphics[width=8.5cm]{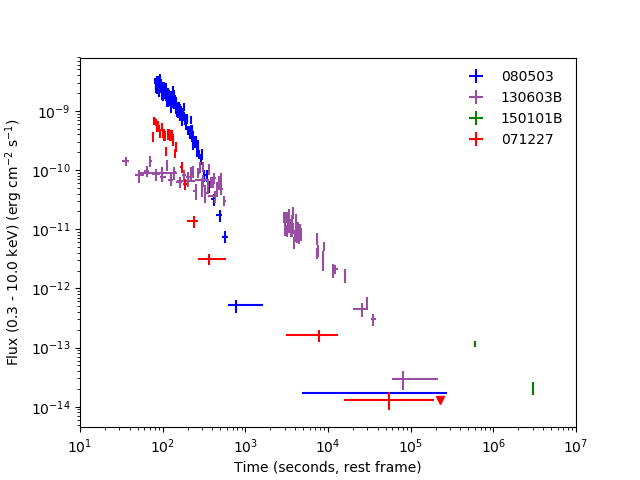}
\caption{The available XRT or \textit{Chandra} light curves for a subsample of SGRBs with kilonova candidates and GRB 071227. In the case of GRB 080503, we used the observed times.}
\label{fig:XRTlc_otherkn}
\end{figure}

\begin{figure}
\includegraphics[width=8.5cm]{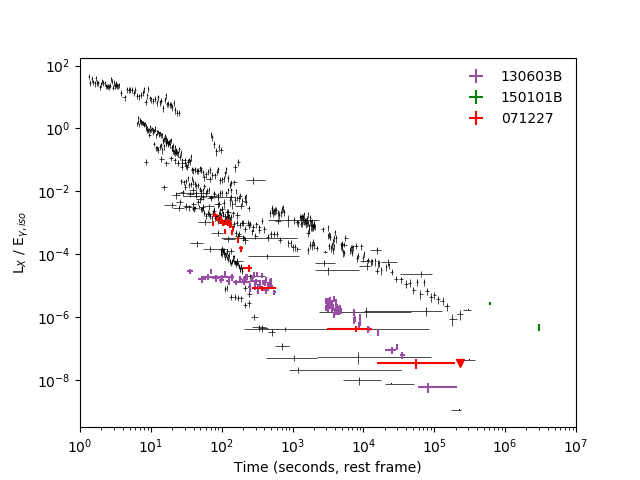}
\caption{The available XRT or \textit{Chandra} light curves normalised to their $E_{\gamma,\text{iso}}$ for the subsample of SGRBs with known redshift given in \citet{DAvanzo14}'s Table 3. We have highlighted the light curves of our kilonova candidate sample and GRB 071227.}
\label{fig:XRTlc_batnormalised}
\end{figure}

We replotted the available XRT light curves for this subsample and compare them more directly with that of GRB 071227 in Figure~\ref{fig:XRTlc_otherkn}. As the XRT light curve for GRB 150101B is contaminated by an AGN, we have used the \textit{Chandra X-ray Observatory} (\textit{Chandra}) light curve from \citet{Fong16}. We find that GRB 071227 is again relatively underluminous at late times, while at early times is also dimmer than the well sampled light curve of GRB 080503. However, GRB 071227 is not atypical for this subsample and is a strong behavioural match to GRB 080503. We have also plotted the light curves of the subsample of SGRBs with known redshift given in Table 3 of \citet{DAvanzo14} normalised to their $E_{\gamma,\text{iso}}$ as in Figure 2b of \citet{Troja18b}, shown in our Figure~\ref{fig:XRTlc_batnormalised}. We find, once again, that GRB 071227 is typical for this sample, although a lack of late time X-ray data means we cannot preclude the same behaviour found in GRB 150101B by \citeauthor{Troja18b}. There is also a clearer agreement between GRB 071227 and GRB 130603B at later times, although GRB 130603B has an early plateau not found in GRB 071227. This analysis of the X-ray agrees with the conclusion of \citet{NicuesaGuelbenzu14} that GRB 071227 is indeed an SGRB.

\begin{figure}
\includegraphics[width=8.5cm]{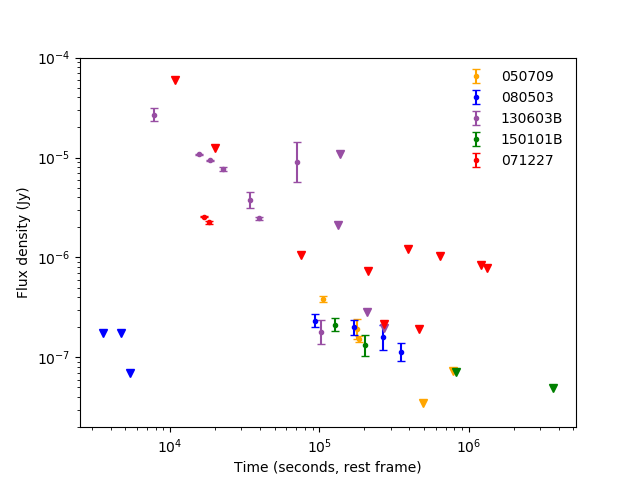}
\caption{The available \textit{r}/\textit{R}-band light curves for our sample of kilonova candidates extrapolated to the same redshift as GRB 071227 without k-corrections and GRB 071227. We have applied an extinction of $E(B-V) = 0.43$ mag to the light curve for GRB 071227 as the other GRBs are little affected by extinction. In the case of GRB 080503, where no redshift could be identified, we use the observed times and magnitudes. The photometry is taken from \citet{Hjorth05b} and \citet{Jin16}; \citet{Perley09}; \citet{Cucchiara13}, \citet{deUgartePostigo14} and \citet{Pandey19}; and \citet{Fong16} for GRBs 050709, 080503, 130603B and 150101B respectively.}
\label{fig:rband_comp}
\end{figure}

Several of our sample GRBs were also observed at optical wavelengths. In Figure~\ref{fig:rband_comp}, we plot the available \textit{r}/\textit{R}-band light curves extrapolated to the same redshift as GRB 071227. It should be noted that GRB 130603B was assumed to be afterglow dominated at optical wavelengths and indications of a kilonova were identified in the NIR. We find that, while no strong conclusions can truly be drawn due to the poor sampling of the light curve, GRB 071227 is qualitatively similar to both GRB 050709 and GRB 150101B. In both these cases, the optical observations were used to infer the presence of a kilonova. We cannot preclude the possibility that GRB 071227 is also consistent with the behaviour exhibited by GRB 080503, where later brightening in the optical has been interpreted as a kilonova with a central engine \citep{Metzger14b}. However, the redshift of GRB 080503 is unknown and unlikely to be low. We also compare the light curve to the models of the AT 2017gfo emission in the same bands from \citet{Gompertz18}, as shown in Figure~\ref{fig:lcplot}, and find them to be generally inconsistent with GRB 071227. However, as the light curve models are strictly phenomonological in regards to AT 2017gfo, this does not preclude other kilonova models being consistent with GRB 071227.

Unfortunately, although the light curves for our subsample of potential kilonovae are often better sampled than GRB 071227, there is a lack of multiband or spectroscopic observations. In particular, it is impossible to evaluate the SEDs of many of these other candidates in order to more directly compare with the transient identified here.

\subsubsection{Possible contributions from other afterglow mechanisms}

There are several other potential mechanisms which could contribute to producing an optical flash. These include a reverse shock passing through the shell as the outflow decelerates \citep{Kobayashi00a} or a hypothetical coupled optical and X-ray flare during the $\sim1.1$ day time period over which our X-ray spectrum was taken.

Short GRBs are typically described by the thin shell regime where the dimensionless parameter $\xi_0>1$ and a reverse shock is inefficient at slowing the shell down. To test if the excess emission at $\sim0.26$ days is the result of a reverse shock, we can estimate the expected flux from the reverse shock given the observed GRB parameters. The maximum flux and the characteristic frequency for the reverse shock are given by \citet{Harrison13}:
\begin{equation}
    F_{\nu,{\rm max},R} \sim F_{\nu,{\rm max},F} C_{\rm F} \Gamma_0 R_{\rm B}^{1/2}  
\end{equation}
and
\begin{equation}
    \nu_{m,R} \sim \nu_{m,F} C_{\rm M} \Gamma_0^{-2} R_{\rm B}^{1/2}    
\end{equation}
where $F_{\nu,{\rm max}}$ is the maximum synchrotron flux and the subscript $R$ or $F$ indicates reverse and forward shock respectively, $\Gamma_0$ is the bulk Lorentz factor of the outflow, $C_{\rm F}$ and $C_{\rm M}$ are correction factors, and $R_{\rm B}$ is a parameter that can be used to boost the magnetic equipartion value $\varepsilon_B$ in the reverse shock region where magnetic fields from the central engine may still contribute.

Assuming an efficiency of $\eta=0.1$ for the $\gamma$-ray energy, the kinetic energy $E_{\rm K}$ of the blastwave is $\sim1.8\times10^{51}$ erg. The dimensionless parameter for this burst is then $\xi_0\approx (l/c T_{90})^{1/2}\Gamma_0^{-4/3}$ where $\Gamma_0 = 100$, the Sedov length is $l=(3E_{\rm K}/4\pi m_p n c^2)^{1/3}$, $n\sim10^{-3}$ is the ambient number density and $m_p$ is the mass of a proton. For these parameters, $\xi_0\sim24$ which confirms the thin nature of the shell. From \citet{Harrison13}, the correction factors are given by $C_{\rm F}\sim(1.5+5\xi_0^{-1.3})^{-1}\sim0.6$ and $C_{\rm M}\sim 5\times10^{-3}+\xi_0^{-3}\sim5\times10^{-3}$, and $F_{\nu,{\rm max},R}\sim 60 F_{\nu,{\rm max},F}$ and $\nu_{m,R}\sim5\times10^{-3} \nu_{m,F}$. The characteristic frequency for the forward shock at the deceleration time, when the reverse shock peaks, is $\nu_{m,F}\sim 5.3\times10^{14}$ Hz where we assume $\varepsilon_B=0.01$, $\varepsilon_e=0.1$ and $\Gamma_0=100$. This indicates that the forward shock in the optical frequencies will be $\sim F_{\nu,{\rm max},F}$ at peak, and the characteristic frequency for the reverse shock will be $\nu_{m,R}\sim2.65\times10^8$ Hz. As $\nu_{m,R}<\nu$, the flux in the optical at the peak of the reverse shock is $F_\nu\sim F_{\nu,{\rm max},R} (\nu/\nu_{m,R})^{-(p-1)/2}\sim 4\times10^{-2} F_{\nu,{\rm max},F}$ and well below the level of the forward shock. Even for a very large $R_{\rm B}$ parameter the reverse shock will be only marginally brighter than the forward shock at deceleration and then decline rapidly as $t^{-(27p+7)/35}$ \citep{Kobayashi00b}.

The peak of the afterglow at frequencies $\nu\geq\nu_m$ is at the deceleration time, typically given by $t_d\sim C_{\rm t}l/c\Gamma_0^{8/3}$, where $C_{\rm t}\sim0.2$ is the correction factor for our parameters. For the outflow to decelerate at $\sim0.26$ days then the Lorentz factor for the outflow would be $\Gamma_0\sim20$; such a low Lorentz factor is inconsistent with a bright GRB \citep{Lamb16} and would reduce the reverse shock flux amplitude. 

The causes of X-ray and optical flares in SGRBs are not yet understood but suggested mechanisms include fragmentation in the accretion disk surrounding the merger remnant and associated variation of the central engine \citep{Perna06} or the fallback of material on eccentric orbits \citep{Rosswog07}. While a flare is unlikely to be a dominant factor in the X-ray spectrum over this period, a flare coinciding with our optical detections could potentially cause both the excess optical flux and steepness. However, there is the suggestion of corresponding features in both the X-ray and optical/NIR light curves of some events \citep[e.g.][]{Malesani07,DAvanzo15}. In particular, \citeauthor{Malesani07} conclude that their optical observations following GRB 050724 are linked to a coincident X-ray flare, and that the rapid fading of their optical light curve is linked to the end of this flare. They do find the SED to be consistent between the X-ray and optical, which is not the case for GRB 071227 although, as previously mentioned, the long time period used to derive our X-ray spectrum could obscure a steepening in the X-ray at a coincident time to our optical detections. It should also be noted, however, that GRB 050724 was an unusual event and evidence for the coupling of optical and X-ray flares is scant.

GRB 071227 has previously been examined for X-ray flares by \citet{Margutti11}, and a potential flare was identified beginning at $\sim150$s, concluding well before our spectral time period and much earlier than our optical detections. No further flares were identified by \citeauthor{Margutti11} and we find no evidence for one in the XRT light curve, a conclusion also reached by \citet{Bernardini11}. Despite this, the low X-ray count rate at later times means we cannot conclusively rule out an X-ray flare at a time when our optical data could be affected.

It is unlikely, therefore, that there is any significant contribution to our optical transient from these mechanisms.

\subsection{GRB 071227 as an extended emission GRB}

GRB 071227 has also been interpreted as an EE GRB \citep{Norris10,Lien16,Gibson17}. With $T_{90} = 142.5 \pm 48.4$~s and $E_{\gamma ,\rm iso} = 1.96^{+0.30}_{-0.29} \times 10^{50}$~erg in the \emph{Swift}-BAT bandpass \citep[k-corrected, cf.][]{Bloom01}, GRB 071227 is one of the longest but least energetic EE GRBs. However, the spread in durations and energies in the population is very limited \citep[e.g.][]{Gompertz13}; hence the large uncertainty on these measurements means it could be of more typical duration and energy. When corrected for redshift, $T_{90, \rm rest} = 102.9 \pm 34.9$~s, which is again both the longest of any EE burst but consistent within errors with being more typical. 29\% of the total energy release occurs within the first 2~s after trigger - one of the highest prompt-to-EE energy ratios in the sample.

\begin{figure}
\includegraphics[width=8.5cm]{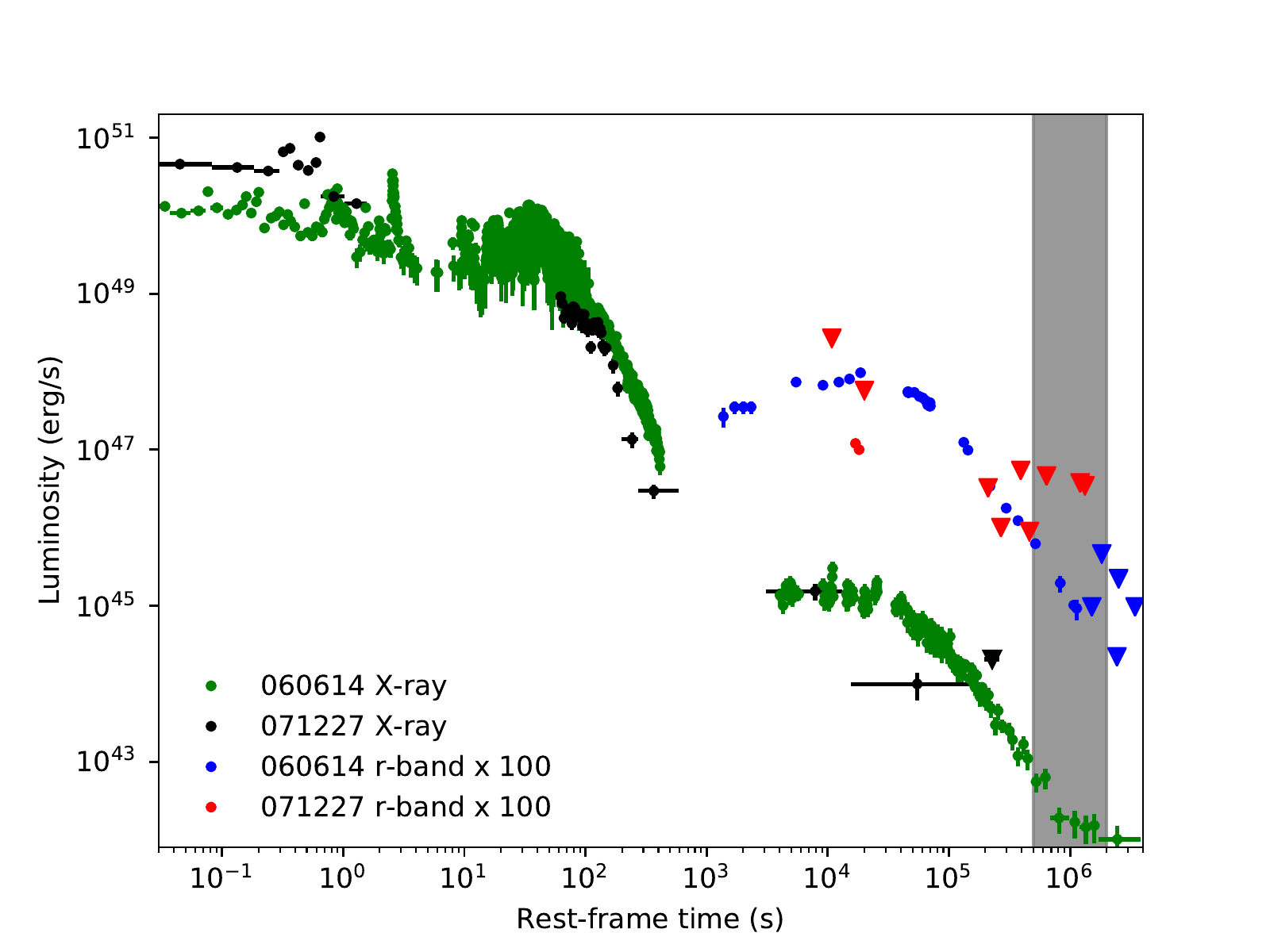}
\caption{GRB 071227 vs GRB 060614, a short GRB with EE and a suspected kilonova. A composite BAT + XRT bolometric light curve is shown in black (green) for GRB 071227 (GRB 060614). The \textit{r}-band light curve of GRB 071227 (GRB 060614) is plotted in red (blue). This has not been k-corrected, but the effect is small, and the optical light curves have also been shifted up by a factor of 100. The grey region marks the time of the excess in the F814W band that was claimed as a possible kilonova in GRB 060614 \citep{Yang15}. \label{fig:EE-comp}}
\end{figure}

One of the EE GRB sample, GRB 060614, contains a suspected kilonova \citep{Xu09,Yang15}. It should be noted that the classification of an EE GRB is still somewhat unclear and some studies take GRB 060614 to be a long burst \citep{Gehrels06,McBreen08,Tanga18}. The EE of GRB 060614 contained a factor of $\sim3$ times more energy than GRB 071227, but the prompt spike energy was comparable between the two. The light curves of both GRBs are shown in Figure~\ref{fig:EE-comp}. After the cessation of EE, the two bursts show markedly different afterglows. GRB 060614 exhibits a long X-ray plateau, symptomatic of energy injection into the emission site, while GRB 071227 appears to decay as a simple power law. However, due to the sparse data, an injection plateau similar to, but shorter than, GRB 060614 cannot be totally discounted. In the \textit{r}-band, GRB 060614 is almost an order of magnitude brighter than that observed in the first epoch of our observations of GRB 071227, with no extinction corrections applied. The light curve of GRB 060614 has been shown to be consistent with a synchrotron afterglow in which the peak frequency is passing through the \textit{r}-band \citep{Gompertz15}. Similarly to GRB 071227, any possible thermal features at optical wavelengths are mostly masked by bright synchrotron emission. The difference in afterglow luminosity between the two bursts could be due to energy injection, as evidenced by the X-ray injection plateau in GRB 060614, or due to a denser circumburst environment around GRB 060614, which would result in a greater flux at both optical and X-ray frequencies. At the time of the F814W filter excess in GRB 060614 \citep{Yang15}, the observations available for GRB 071227 are not very constraining; the 3 $\sigma$ upper limits in the \textit{r} and \textit{i}-bands permit an optical afterglow or kilonova up to the luminosity of the one seen in GRB 060614.

\section{Host galaxy properties}

\subsection{Host galaxy morphology and GRB offset}
\label{sec:HostMorphology}

The host galaxy of GRB 071227 can be clearly identified as a late type spiral with an observed optical disk radius of $\sim$20 kpc in both the data presented here, and in \citet{DAvanzo09} and \citet{Fong13a}. The disk is viewed mostly edge on, with an inclination of $\sim$73\degr~which we derive from fitting elliptical isophotes to our highest quality image, the second epoch \textit{r}-band image. The host has a rest frame {\em B}-band absolute magnitude of $M_B=-19.3$ \citep{DAvanzo09}. \citet{Fong13b}  examine the morphology of the galaxy and fit the surface brightness profile to a S\'{e}rsic model consistent with an exponential disk.

\citet{Leibler10}, meanwhile, fit SEDs to optical and infrared observations of the host to identify its age and mass. Fitting to a single stellar population indicates an age of 0.49 Gyr and a mass of $0.025\times 10^{12}$ M$_{\sun}$. A fit using only the $K$-band fluxes gave a maximum stellar mass of $0.25\times 10^{12}$ M$_{\sun}$.

Radio observations of the host have also been performed \citep{NicuesaGuelbenzu14}, detecting a total integrated flux density of $F_{\nu} = 43 \pm 11$ $\mu$Jy at 5.5 GHz. While the centre of the radio emission is offset from the galactic bulge visible in the optical, this is deemed insignificant and within error bounds. \citeauthor{NicuesaGuelbenzu14} conclude that the galaxy is inactive, due to its position on a \textit{WISE} 3.4-4.6-12 $\mu$m (W1-W2-W3) colour-colour plot (using Vega magnitudes) and morphology at 5.5 GHz. This colour-colour plot indicates the host to be a LIRG (luminous infrared galaxy). SED fitting with additional data from \citet{Leibler10} and the \textit{WISE} survey indicates a stellar mass of $0.32\times 10^{12}$ M$_{\sun}$.

We used the position of the transient identified from our subtractions to measure an offset of 2.91 $\pm$ 0.10 arcsec from the galactic centre. At $z = 0.381$, this corresponds to 15.71 $\pm$ 0.54 kpc, consistent with the $\sim 15$ kpc offsets found by \citeauthor{DAvanzo09} and \citeauthor{Fong13b}.

\subsection{Spatially resolved emission line spectroscopy}
\label{sec:Spectroscopy}

The 11 subspectra extracted from our GMOS-S spectroscopy were separately analysed to determine spatially-resolved properties of the stars and gas across the host galaxy, although it should be noted that the subspectra are not truly independent. As well as analysing the individual subspectra we split the galaxy into two disk sections (subspectra 1--4 and 8--11) and a central region (subspectra 5--7) in order to increase the signal to noise ratio (SNR) of the spectra being analysed. Figure~\ref{fig:apperture} demonstrates this. This analysis was done following the procedure explained in \citet{Lyman18b}. Briefly, the stellar continuum was fit using {\sc starlight} \citep{CidFernandes05} and the resulting emission-line spectrum (after subtraction of this stellar continuum model) was fit with a series of Gaussians centered at locations of strong nebular lines. The emission line ratios and fluxes were used to derive properties such as extinction, via the Balmer decrement, and metallicity, using indicators from \citet{Pettini04}.

Many of the individual subspectra had too low SNR in order to robustly fit either the stellar continuum or the nebular lines. Even with binning of the subspectra, the SNR was not sufficient to tightly constrain the fit of the stellar continuum, which are notoriously prone to multiple degeneracies and prove problematic to interpret even in high SNR data. For the nucleus, where the continuum signal is strongest, we find the best-fitting stellar population model to be dominated by an old stellar population at $\gtrsim$10~Gyr with a modest (few percent by mass) contribution from younger populations at $\lesssim$1~Gyr. This is somewhat consistent with the young+old simple stellar population model used by \citet{Leibler10}, a model fitted to the galaxy's observed SED in which the stellar population is divided into old stars with an age identical to that of the universe at the host's redshift, and a more recently formed stellar population with a fitted age of 0.36 Gyr. They inferred stellar masses of $<0.100\times 10^{12}$ M$_{\sun}$ for the old population and $0.020\times 10^{12}$ M$_{\sun}$ for the young.

At the location of the GRB (subspectrum 2), our emission line fits gave $E(B-V)_\text{gas} = 0.54\pm{0.37}$ mag of extinction based on the Balmer decrement, assuming an intrinsic $F(H\alpha)/F(H\beta) = 2.86$ \citep{Osterbrock06}, as expected from our spectral fits of GRB 071227. When analysing the combined disk spectrum from subspectra 1--4 to obtain a higher SNR in the lines, we obtain $E(B-V)_\text{gas} = 0.53\pm{0.16}$ mag, greater extinction than identified in many other SGRB hosts examined \citep{Savaglio09,Yoshida19}. The metallicity at the GRB location was found to be $12 + \log(\text{O}/\text{H}) = 8.5 \pm 0.3$ and $8.5 \pm 0.2$ dex, using the N2 and O3N2 relations of \citet{Pettini04}, respectively. When considering the summed disk spectrum on the GRB side of the galaxy, these values are $8.7 \pm 0.1$ and $8.6 \pm 0.1$ dex respectively. These are consistent with the range of 8.2 to 8.8 inferred by \citet{DAvanzo09} for this host, and are also comparable to the sample of local galaxies \citep[$\sim7.5 - 9.2$;][]{Walter08} and the Milky Way \citep[$\sim$8.7;][]{Baumgartner06}. Of particular note is the similarity to the metallicity inferred for the host of SGRB 080905A \citep[$\sim8.4 - 8.8$;][]{Rowlinson10}. Note that these gas phase metallicities are relevant only for the ongoing star formation and young stellar population of the host, and thus may not be indicative of the metallicity of a possible very old progenitor. When analysing the disk on the opposite side of the nucleus from the GRB, we find very similar metallicity values but a somewhat higher $E(B-V)_\text{gas} \sim1.4$ mag. The difference in extinction on opposing sides of the disk may be a result of viewing the host edge on and the geometry of the dust lanes with respect to the spiral arms.

We measure a total host unobscured \zt\ luminosity of $\sim5 \times 10^{40}$ erg s$^{-1}$, which, using equation 3 from \citet{Kennicutt98}, gives a star formation rate of $0.7 \pm 0.2$ M$_{\sun}$\,yr$^{-1}$. Alternatively, using equation 10 from \citet{Kewley04} and our mean metallicity of 8.6, gives a rate of $0.2^{+0.4}_{-0.2}$ M$_{\sun}$\,yr$^{-1}$. \citet{DAvanzo09} used their \zt\ luminosity of $3.9 \times 10^{40}$ erg s$^{-1}$ to identify a star formation rate of 0.6 M$_{\sun}$\,yr$^{-1}$, consistent with our calculations. Returning to the radio observations performed by \citet{NicuesaGuelbenzu14}, a fit using the \textsc{grasil} software indicates a star formation rate of 24 M$_{\sun}$\,yr$^{-1}$ implying the host is undergoing an intensive star forming period, however. The discrepancy with the optical emission line diagnostics is attributed to large amounts of optically-obscured star formation by \citeauthor{NicuesaGuelbenzu14}.

\subsection{Rotation curve}
\label{sec:RotCurve}

The nearly edge-on orientation of the host galaxy gives us an excellent view on the rotation curve of the galaxy. In the 2D spectrum a clear slant can be seen in the \st, \ha, \nt, \zd, \hb\ and \zt\ lines.  To determine the rotation curve of the galaxy we used the \textsc{ngaussfit} routines in the {\sc IRAF} {\sc stsdas} package. We extracted postage stamp spectra using a 5 pixels bin in the spatial direction (corresponding to 0.73 arcseconds or 3.79 kpc) and 150 pixels in the dispersion direction from the 2D spectra. We then fitted the continuum level and slope, and fitted the line with a Gaussian profile to determine its centre position. The postage stamps were shifted with 3 pixels in spatial direction after each fit. This minimizes the effect of weak cosmic rays and residuals of skylines, but also results in nonindependent datapoints.

The \ha\ line is bright, but the Balmer stellar atmospheric absorption is strong near the nucleus; we did not explicitly fit for the absorption in the line centroid measurements. 

\begin{figure}
\includegraphics[width=\columnwidth]{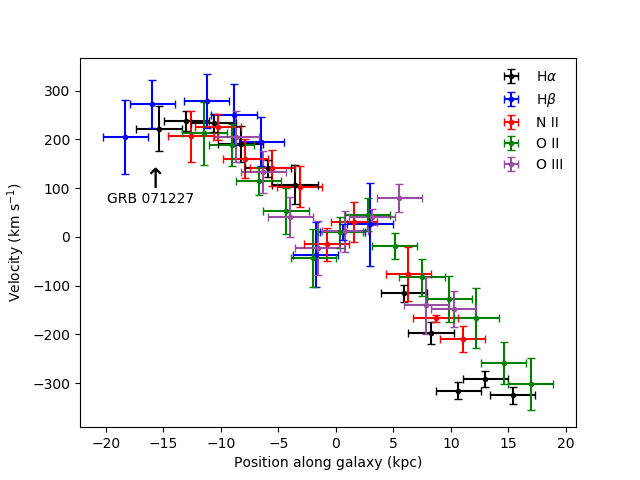}
\caption{The rotation curve of the host galaxy, velocities are as observed (not corrected for the inclination). The position of the GRB  in the South-East tip of the galaxy moving away from us is noted.}
\label{fig:rv}
\end{figure}

The \zt\ doublet is not as bright as the \ha\ line, but the local continuum is extremely weak, making the line center easy to fit, although it is more difficult to determine the galaxy centre pixel. Furthermore, the \ha\ and \zt\ lines are located at the extreme ends of the spectrum, where the arc spectrum has only a few lines, making the wavelength calibration less certain. We therefore also fitted the \nt, \zd\ and \hb\ lines, but did not use the \st\ doublet which is too close to the end of the chip. The resulting radial velocity measurements, including only points with a velocity uncertainty better than 100 km\,s$^{-1}$, are shown in Figure~\ref{fig:rv}. The GRB took place in the approaching, South-East, end of the galaxy as indicated.

The rotation curve appears to consist of a disk-like profile with a significant undulation around the centre, particularly clear in \zt, \zd\ and \nt. This undulation is possibly caused by a centrally located bar structure, a large spiral arm or a counterrotating disk, but to establish this definitively, a much higher signal to noise is required so the line centres can be measured in smaller bins and the zero position and velocity can be more accurately determined. We can, however, estimate the dynamical mass of the galaxy through the velocity at which the rotation curve levels off, which we evaluate by using a weighted mean of a rebinned dataset, using inverse variance weighting taking into account the bin overlap. Following their initial kicks, the mass of their host is an important parameter in the travel of a compact binary. Identifying the host's mass, therefore, is useful for understanding and constraining the potential offsets of SGRBs from their hosts, which could enable host associations and therefore redshifts in other SGRBs. We correct for the inclination of $\sim$73\degr\ as identified in section~\ref{sec:HostMorphology}, and find a dynamical mass of $\sim 0.32 \times 10^{12}$ M$_{\sun}$, consistent with the stellar masses of $0.32\times 10^{12}$ M$_{\sun}$ found by \citet{NicuesaGuelbenzu14} and $0.25\times 10^{12}$ M$_{\sun}$ found by \citet{Leibler10}, an agreement typical for galaxies of this mass \citep{Drory04}. This mass is approximately 20 to 30\% of that inferred for the Milky Way \citep{McMillan17} and an order of magnitude larger than that of the host of GRB 080905A \citep{Rowlinson10}.

\subsection{Host environment}

The evolution of a galaxy, and by extension, the objects within, can be strongly influenced by its environment. It is therefore important to identify key features of a host galaxy's environment, such as clustering, to better inform future observations of GRBs. In addition, determining the relationship between SGRB hosts and their environments could allow redshifts to be inferred for hostless SGRBs.

\begin{figure}
\centerline{\includegraphics[width=8.5cm]{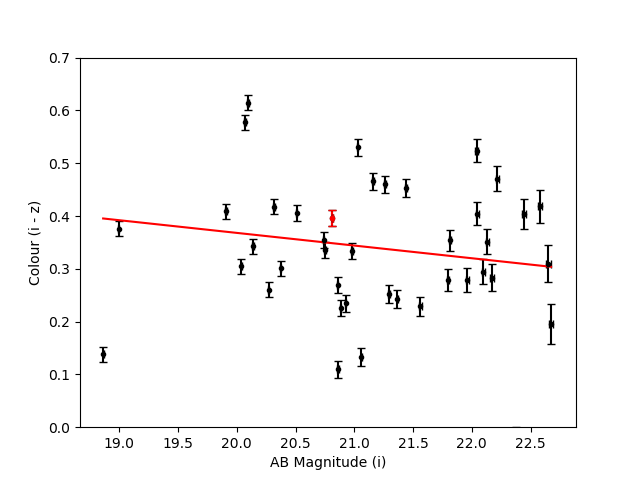}}
\caption{A colour-magnitude plot and the expected red sequence at $z=0.381$ (red, solid) for the \textit{i}/\textit{z} filter combination. The host of GRB 071227 is highlighted in red.} 
\label{fig:RedSequence}
\end{figure}

Whilst it appears from our imaging data that the host of GRB 071227 is in a local cluster, we can verify this assertion using red sequencing: a colour-magnitude plot of the galaxies within a cluster will display a linear relationship known as the red sequence, the specifics of which are defined by the redshift of the cluster \citep{Gladders00}. We produced a plot of the colour-magnitude relation for each combination of filters for the galaxies visible within the GMOS-S field, an example of which is shown in Figure~\ref{fig:RedSequence}, including the expected red sequences at our redshift of 0.381, which we constructed using the Millenium \textit{N}-body simulation \citep{Springel05} and a similar methodology to \citet{Stott09}. The galaxy evolution in our case is defined in \citet{Lagos12}. Our red sequences consistently indicate that the host of GRB 071227, marked in red in the plots, is indeed part of a local cluster of galaxies.

\section{Conclusions}

We have investigated the short duration GRB 071227 at X-ray and optical wavelengths, and find apparent excess optical flux inconsistent with the inferred afterglow extrapolated from fitting only the X-ray data. Using an MCMC method to fit the SED, however, we find that both the X-ray and optical transients are generally consistent with an afterglow where $\Gamma \sim 2.1-2.2$, although an excess remains in the \textit{z}-band.

We also investigate other possible causes of this excess. We find that fitting the excess to a blackbody is generally inconsistent with the results expected from kilonova models, requiring superluminal motion and a high ejecta mass of $\sim 0.1$ M$_{\sun}$, and varying the afterglow parameters and extinction generally fails to reconcile this. We cannot fully rule out some thermal contribution, however, which could be responsible for the \textit{z}-band excess. Additionally, we have compared GRB 071227 to several kilonova candidates, finding the \textit{r}/\textit{R}-band and XRT lightcurves of GRB 071227 to be consistent with these observations and with the sample of SGRBs in general.

We find the excess is also inconsistent with being the result of a reverse shock and there is no evidence for a coupled X-ray optical flare as has been previously suggested in the case of GRB 050724. We have examined GRB 071227 in the context of EE GRBs, particularly in comparison to GRB 060614.

Finally, we have performed spectroscopic analysis of the host of GRB 071227. We find that the host is a spiral galaxy with a dynamical mass of $\sim0.35$ $\times 10^{12}$ M$_{\sun}$. In addition, we have derived a majority stellar population age of $\ga 10$ Gyr with a small contribution from a younger population, a mean metallicity of $12 + \log(\text{O}/\text{H}) = 8.6$ dex and a star formation rate of 0.2 to 0.7 M$_{\sun}$ yr$^{-1}$. We evaluate the extinction of the galaxy finding it to be dustier than many other SGRB hosts with $E(B-V)_\text{gas} = 0.54\pm{0.37}$ near the GRB and $E(B-V)_\text{gas} \sim1.4$ in the other disk region. We also used red sequencing to show that the galaxy is situated in a local cluster. Overall, despite having a higher intrinsic extinction than most other examples, we find the host of GRB 071227 to be a typical late-type SGRB host, less massive and with comparable metallicity to the Milky Way.

\section*{Acknowledgements}

Based on observations obtained at the Gemini Observatory, which is operated by the Association of Universities for Research in Astronomy, Inc., under a cooperative agreement with the NSF on behalf of the Gemini partnership: the National Science Foundation (United States), the National Research Council (Canada), CONICYT (Chile), Ministerio de Ciencia, Tecnolog\'{i}a e Innovaci\'{o}n Productiva (Argentina), and Minist\'{e}rio da Ci\^{e}ncia, Tecnologia e Inova\c{c}\~{a}o (Brazil) (acquired through the Gemini Observatory Archive and processed using the Gemini IRAF package).

This research has made use of the services of the ESO Science Archive Facility. Based on observations collected at the European Southern Observatory under ESO programme 080.D-0906(G).

This work made use of data supplied by the UK \textit{Swift} Science Data Centre at the University of Leicester.

The national facility capability for SkyMapper has been funded through ARC LIEF grant LE130100104 from the Australian Research Council, awarded to the University of Sydney, the Australian National University, Swinburne University of Technology, the University of Queensland, the University of Western Australia, the University of Melbourne, Curtin University of Technology, Monash University and the Australian Astronomical Observatory. SkyMapper is owned and operated by The Australian National University's Research School of Astronomy and Astrophysics. The survey data were processed and provided by the SkyMapper Team at ANU. The SkyMapper node of the All-Sky Virtual Observatory (ASVO) is hosted at the National Computational Infrastructure (NCI). Development and support the SkyMapper node of the ASVO has been funded in part by Astronomy Australia Limited (AAL) and the Australian Government through the Commonwealth's Education Investment Fund (EIF) and National Collaborative Research Infrastructure Strategy (NCRIS), particularly the National eResearch Collaboration Tools and Resources (NeCTAR) and the Australian National Data Service Projects (ANDS).

RAJE, PTO, RLCS, GPL and NRT acknowledge funding from the Science and Technology Facilities Council.

KW and BPG have received funding from the European Research Council (ERC) under the European Union's Horizon 2020 research and innovation programme (grant agreement no 725246, TEDE, PI Levan).

SR has been supported by the Swedish Research Council (VR) under grant number 2016-03657 3, by the Swedish National Space Board under grant number Dnr. 107/16 and by the research environment grant ``Gravitational Radiation and Electromagnetic Astrophysical Transients'' (GREAT) funded by the Swedish Research council (VR) under Dnr. 2016-06012. Support from the COST Actions on neutron stars (PHAROS; CA16214) and black holes and gravitational waves (GWerse; CA16104) are gratefully acknowledged. 

The authors would like to thank Andy Fruchter and John Graham for their contributions to the Gemini South observations. We also thank Brian Metzger and Masaomi Tanaka for helpful discussion and suggestions. Finally, we thank the anonymous referee for their useful comments.




\bibliographystyle{mnras}
\bibliography{071227,software}

\begin{thebibliography}{}
\makeatletter
\relax
\def\mn@urlcharsother{\let\do\@makeother \do\$\do\&\do\#\do\^\do\_\do\%\do\~}
\def\mn@doi{\begingroup\mn@urlcharsother \@ifnextchar [ {\mn@doi@}
  {\mn@doi@[]}}
\def\mn@doi@[#1]#2{\def\@tempa{#1}\ifx\@tempa\@empty \href
  {http://dx.doi.org/#2} {doi:#2}\else \href {http://dx.doi.org/#2} {#1}\fi
  \endgroup}
\def\mn@eprint#1#2{\mn@eprint@#1:#2::\@nil}
\def\mn@eprint@arXiv#1{\href {http://arxiv.org/abs/#1} {{\tt arXiv:#1}}}
\def\mn@eprint@dblp#1{\href {http://dblp.uni-trier.de/rec/bibtex/#1.xml}
  {dblp:#1}}
\def\mn@eprint@#1:#2:#3:#4\@nil{\def\@tempa {#1}\def\@tempb {#2}\def\@tempc
  {#3}\ifx \@tempc \@empty \let \@tempc \@tempb \let \@tempb \@tempa \fi \ifx
  \@tempb \@empty \def\@tempb {arXiv}\fi \@ifundefined
  {mn@eprint@\@tempb}{\@tempb:\@tempc}{\expandafter \expandafter \csname
  mn@eprint@\@tempb\endcsname \expandafter{\@tempc}}}

\bibitem[\protect\citeauthoryear{{Abbott} et~al.}{{Abbott}
  et~al.}{2017}]{Abbott17}
{Abbott} B.~P.,  et~al., 2017, \mn@doi [\apj] {10.3847/2041-8213/aa91c9}, \href
  {https://ui.adsabs.harvard.edu/#abs/2017ApJ...848L..12A} {848}

\bibitem[\protect\citeauthoryear{{Alard}}{{Alard}}{2000}]{Alard00}
{Alard} C.,  2000, \mn@doi [\aaps] {10.1051/aas:2000214}, \href
  {http://adsabs.harvard.edu/abs/2000A%26AS..144..363A} {144, 363}

\bibitem[\protect\citeauthoryear{{Arnaud}}{{Arnaud}}{1996}]{Arnaud96}
{Arnaud} K.~A.,  1996, in {Jacoby} G.~H.,  {Barnes} J.,  eds,  Astronomical
  Society of the Pacific Conference Series Vol. 101, Astronomical Data Analysis
  Software and Systems V. p.~17

\bibitem[\protect\citeauthoryear{{Ascenzi} et~al.,}{{Ascenzi}
  et~al.}{2019}]{Ascenzi19}
{Ascenzi} S.,  et~al., 2019, \mn@doi [\mnras] {10.1093/mnras/stz891}, \href
  {https://ui.adsabs.harvard.edu/abs/2019MNRAS.486..672A} {486, 672}

\bibitem[\protect\citeauthoryear{{Barnes}, {Kasen}, {Wu}  \&
  {Mart{\'{\i}}nez-Pinedo}}{{Barnes} et~al.}{2016}]{Barnes16}
{Barnes} J.,  {Kasen} D.,  {Wu} M.-R.,   {Mart{\'{\i}}nez-Pinedo} G.,  2016,
  \mn@doi [\apj] {10.3847/0004-637X/829/2/110}, \href
  {http://adsabs.harvard.edu/abs/2016ApJ...829..110B} {829, 110}

\bibitem[\protect\citeauthoryear{{Baumgartner} \& {Mushotzky}}{{Baumgartner} \&
  {Mushotzky}}{2006}]{Baumgartner06}
{Baumgartner} W.~H.,  {Mushotzky} R.~F.,  2006, \mn@doi [\apj]
  {10.1086/499619}, \href
  {https://ui.adsabs.harvard.edu/\#abs/2006ApJ...639..929B} {639, 929}

\bibitem[\protect\citeauthoryear{{Beardmore}, {Page}  \&
  {Sakamoto}}{{Beardmore} et~al.}{2007}]{Beardmore07}
{Beardmore} A.~P.,  {Page} K.~L.,   {Sakamoto} T.,  2007, GRB Coordinates
  Network, \href {http://adsabs.harvard.edu/abs/2007GCN..7153....1B} {7153}

\bibitem[\protect\citeauthoryear{{Becker}}{{Becker}}{2015}]{Becker15}
{Becker} A.,  2015, {HOTPANTS: High Order Transform of PSF ANd Template
  Subtraction}, Astrophysics Source Code Library (\mn@eprint {ascl} {1504.004})

\bibitem[\protect\citeauthoryear{{Berger}, {Morrell}  \& {Roth}}{{Berger}
  et~al.}{2007}]{Berger07a}
{Berger} E.,  {Morrell} N.,   {Roth} M.,  2007, GRB Coordinates Network, \href
  {http://adsabs.harvard.edu/abs/2007GCN..7151....1B} {7151}

\bibitem[\protect\citeauthoryear{{Bernardini}, {Margutti}, {Chincarini},
  {Guidorzi}  \& {Mao}}{{Bernardini} et~al.}{2011}]{Bernardini11}
{Bernardini} M.~G.,  {Margutti} R.,  {Chincarini} G.,  {Guidorzi} C.,   {Mao}
  J.,  2011, \mn@doi [\aap] {10.1051/0004-6361/201015703}, \href
  {https://ui.adsabs.harvard.edu/#abs/2011A&A...526A..27B} {526, A27}

\bibitem[\protect\citeauthoryear{{Bertin} \& {Arnouts}}{{Bertin} \&
  {Arnouts}}{1996}]{SExtractor96}
{Bertin} E.,  {Arnouts} S.,  1996, \mn@doi [\aaps] {10.1051/aas:1996164}, \href
  {{http://adsabs.harvard.edu/abs/1996A%26AS..117..393B}} {117, 393}

\bibitem[\protect\citeauthoryear{{Bloom}, {Frail}  \& {Sari}}{{Bloom}
  et~al.}{2001}]{Bloom01}
{Bloom} J.~S.,  {Frail} D.~A.,   {Sari} R.,  2001, \mn@doi [\aj]
  {10.1086/321093}, \href {http://adsabs.harvard.edu/abs/2001AJ....121.2879B}
  {121, 2879}

\bibitem[\protect\citeauthoryear{{Brege} et~al.,}{{Brege}
  et~al.}{2018}]{Brege18}
{Brege} W.,  et~al., 2018, \mn@doi [\prd] {10.1103/PhysRevD.98.063009}, \href
  {http://cdsads.u-strasbg.fr/abs/2018PhRvD..98f3009B} {98, 063009}

\bibitem[\protect\citeauthoryear{{Cardelli}, {Clayton}  \& {Mathis}}{{Cardelli}
  et~al.}{1989}]{Cardelli89}
{Cardelli} J.~A.,  {Clayton} G.~C.,   {Mathis} J.~S.,  1989, \mn@doi [\apj]
  {10.1086/167900}, \href {http://adsabs.harvard.edu/abs/1989ApJ...345..245C}
  {345, 245}

\bibitem[\protect\citeauthoryear{{Chevalier} \& {Li}}{{Chevalier} \&
  {Li}}{1999}]{Chevalier99}
{Chevalier} R.~A.,  {Li} Z.-Y.,  1999, \mn@doi [\apj] {10.1086/312147}, \href
  {https://ui.adsabs.harvard.edu/#abs/1999ApJ...520L..29C} {520, L29}

\bibitem[\protect\citeauthoryear{{Chevalier} \& {Li}}{{Chevalier} \&
  {Li}}{2000}]{Chevalier00}
{Chevalier} R.~A.,  {Li} Z.-Y.,  2000, \mn@doi [\apj] {10.1086/308914}, \href
  {https://ui.adsabs.harvard.edu/#abs/2000ApJ...536..195C} {536, 195}

\bibitem[\protect\citeauthoryear{{Cid Fernandes}, {Mateus}, {Sodr{\'e}},
  {Stasi{\'n}ska}  \& {Gomes}}{{Cid Fernandes} et~al.}{2005}]{CidFernandes05}
{Cid Fernandes} R.,  {Mateus} A.,  {Sodr{\'e}} L.,  {Stasi{\'n}ska} G.,
  {Gomes} J.~M.,  2005, \mn@doi [\mnras] {10.1111/j.1365-2966.2005.08752.x},
  \href {http://adsabs.harvard.edu/abs/2005MNRAS.358..363C} {358, 363}

\bibitem[\protect\citeauthoryear{{Covino} et~al.,}{{Covino}
  et~al.}{2017}]{Covino17}
{Covino} S.,  et~al., 2017, \mn@doi [Nature Astronomy]
  {10.1038/s41550-017-0285-z}, \href
  {https://ui.adsabs.harvard.edu/\#abs/2017NatAs...1..791C} {1, 791}

\bibitem[\protect\citeauthoryear{{Cowperthwaite} et~al.,}{{Cowperthwaite}
  et~al.}{2017}]{Cowperthwaite17}
{Cowperthwaite} P.~S.,  et~al., 2017, \mn@doi [\apj]
  {10.3847/2041-8213/aa8fc7}, \href
  {https://ui.adsabs.harvard.edu/\#abs/2017ApJ...848L..17C} {848, L17}

\bibitem[\protect\citeauthoryear{{Cucchiara} et~al.,}{{Cucchiara}
  et~al.}{2013}]{Cucchiara13}
{Cucchiara} A.,  et~al., 2013, \mn@doi [\apj] {10.1088/0004-637X/777/2/94},
  \href {https://ui.adsabs.harvard.edu/abs/2013ApJ...777...94C} {777, 94}

\bibitem[\protect\citeauthoryear{{D'Avanzo}}{{D'Avanzo}}{2015}]{DAvanzo15}
{D'Avanzo} P.,  2015, \mn@doi [Journal of High Energy Astrophysics]
  {10.1016/j.jheap.2015.07.002}, \href
  {https://ui.adsabs.harvard.edu/#abs/2015JHEAp...7...73D} {7, 73}

\bibitem[\protect\citeauthoryear{{D'Avanzo} et~al.,}{{D'Avanzo}
  et~al.}{2007}]{DAvanzo07a}
{D'Avanzo} P.,  et~al., 2007, GRB Coordinates Network, \href
  {http://adsabs.harvard.edu/abs/2007GCN..7149....1D} {7149}

\bibitem[\protect\citeauthoryear{{D'Avanzo} et~al.,}{{D'Avanzo}
  et~al.}{2009}]{DAvanzo09}
{D'Avanzo} P.,  et~al., 2009, \mn@doi [\aap] {10.1051/0004-6361/200811294},
  \href {http://adsabs.harvard.edu/abs/2009A%26A...498..711D} {498, 711}

\bibitem[\protect\citeauthoryear{{D'Avanzo} et~al.,}{{D'Avanzo}
  et~al.}{2014}]{DAvanzo14}
{D'Avanzo} P.,  et~al., 2014, \mn@doi [\mnras] {10.1093/mnras/stu994}, \href
  {http://adsabs.harvard.edu/abs/2014MNRAS.442.2342D} {442, 2342}

\bibitem[\protect\citeauthoryear{{Dorman} \& {Arnaud}}{{Dorman} \&
  {Arnaud}}{2001}]{Dorman01}
{Dorman} B.,  {Arnaud} K.~A.,  2001, in {Harnden} Jr. F.~R.,  {Primini} F.~A.,
   {Payne} H.~E.,  eds,  Astronomical Society of the Pacific Conference Series
  Vol. 238, Astronomical Data Analysis Software and Systems X. p.~415

\bibitem[\protect\citeauthoryear{{Draper}, {Gray}, {Berry}  \&
  {Taylor}}{{Draper} et~al.}{2014}]{Draper14}
{Draper} P.~W.,  {Gray} N.,  {Berry} D.~S.,   {Taylor} M.,  2014, {GAIA:
  Graphical Astronomy and Image Analysis Tool}, Astrophysics Source Code
  Library (\mn@eprint {ascl} {1403.024})

\bibitem[\protect\citeauthoryear{{Drory}, {Bender}  \& {Hopp}}{{Drory}
  et~al.}{2004}]{Drory04}
{Drory} N.,  {Bender} R.,   {Hopp} U.,  2004, \mn@doi [\apj] {10.1086/426502},
  \href {https://ui.adsabs.harvard.edu/#abs/2004ApJ...616L.103D} {616, L103}

\bibitem[\protect\citeauthoryear{{Duffell}, {Quataert}, {Kasen}  \&
  {Klion}}{{Duffell} et~al.}{2018}]{Duffell18}
{Duffell} P.~C.,  {Quataert} E.,  {Kasen} D.,   {Klion} H.,  2018, \mn@doi
  [\apj] {10.3847/1538-4357/aae084}, \href
  {http://adsabs.harvard.edu/abs/2018ApJ...866....3D} {866, 3}

\bibitem[\protect\citeauthoryear{{Duflo} \& {Zuker}}{{Duflo} \&
  {Zuker}}{1995}]{Duflo95}
{Duflo} J.,  {Zuker} A.~P.,  1995, \mn@doi [\prc] {10.1103/PhysRevC.52.R23},
  \href {https://ui.adsabs.harvard.edu/#abs/1995PhRvC..52...23D} {52, R23}

\bibitem[\protect\citeauthoryear{ESO}{ESO}{2018}]{ESO2018}
ESO 2018, FORS2 User Manual issue 104; VLT-MAN-ESO-13100-1543.
European Southern Observatory

\bibitem[\protect\citeauthoryear{{Evans} et~al.,}{{Evans}
  et~al.}{2007}]{Evans07}
{Evans} P.~A.,  et~al., 2007, \mn@doi [\aap] {10.1051/0004-6361:20077530},
  \href {http://adsabs.harvard.edu/abs/2007A%26A...469..379E} {469, 379}

\bibitem[\protect\citeauthoryear{{Evans} et~al.,}{{Evans}
  et~al.}{2009}]{Evans09}
{Evans} P.~A.,  et~al., 2009, \mn@doi [\mnras]
  {10.1111/j.1365-2966.2009.14913.x}, \href
  {http://adsabs.harvard.edu/abs/2009MNRAS.397.1177E} {397, 1177}

\bibitem[\protect\citeauthoryear{Fong \& Berger}{Fong \&
  Berger}{2013}]{Fong13b}
Fong W.,  Berger E.,  2013, \mn@doi [\apj] {10.1088/0004-637X/776/1/18}, \href
  {http://adsabs.harvard.edu/abs/2013ApJ...776...18F} {776, 18}

\bibitem[\protect\citeauthoryear{{Fong} et~al.,}{{Fong} et~al.}{2013}]{Fong13a}
{Fong} W.,  et~al., 2013, \mn@doi [\apj] {10.1088/0004-637X/769/1/56}, \href
  {http://adsabs.harvard.edu/abs/2013ApJ...769...56F} {769, 56}

\bibitem[\protect\citeauthoryear{{Fong}, {Berger}, {Margutti}  \&
  {Zauderer}}{{Fong} et~al.}{2015}]{Fong15}
{Fong} W.,  {Berger} E.,  {Margutti} R.,   {Zauderer} B.~A.,  2015, \mn@doi
  [\apj] {10.1088/0004-637X/815/2/102}, \href
  {http://adsabs.harvard.edu/abs/2015ApJ...815..102F} {815, 102}

\bibitem[\protect\citeauthoryear{{Fong} et~al.,}{{Fong} et~al.}{2016}]{Fong16}
{Fong} W.,  et~al., 2016, \mn@doi [\apj] {10.3847/1538-4357/833/2/151}, \href
  {http://adsabs.harvard.edu/abs/2016ApJ...833..151F} {833, 151}

\bibitem[\protect\citeauthoryear{{Foucart} et~al.,}{{Foucart}
  et~al.}{2014}]{Foucart14}
{Foucart} F.,  et~al., 2014, \mn@doi [\prd] {10.1103/PhysRevD.90.024026}, \href
  {http://cdsads.u-strasbg.fr/abs/2014PhRvD..90b4026F} {90, 024026}

\bibitem[\protect\citeauthoryear{{Foucart}, {Duez}, {Kidder}, {Nissanke},
  {Pfeiffer}  \& {Scheel}}{{Foucart} et~al.}{2019}]{Foucart19}
{Foucart} F.,  {Duez} M.~D.,  {Kidder} L.~E.,  {Nissanke} S.,  {Pfeiffer}
  H.~P.,   {Scheel} M.~A.,  2019, arXiv e-prints, \href
  {http://cdsads.u-strasbg.fr/abs/2019arXiv190309166F} {}

\bibitem[\protect\citeauthoryear{{Fox} et~al.,}{{Fox} et~al.}{2005}]{Fox05}
{Fox} D.~B.,  et~al., 2005, \mn@doi [\nat] {10.1038/nature04189}, \href
  {http://adsabs.harvard.edu/abs/2005Natur.437..845F} {437, 845}

\bibitem[\protect\citeauthoryear{{Freudling}, {Romaniello}, {Bramich},
  {Ballester}, {Forchi}, {Garc{\'{\i}}a-Dabl{\'o}}, {Moehler}  \&
  {Neeser}}{{Freudling} et~al.}{2013}]{Freudling13}
{Freudling} W.,  {Romaniello} M.,  {Bramich} D.~M.,  {Ballester} P.,  {Forchi}
  V.,  {Garc{\'{\i}}a-Dabl{\'o}} C.~E.,  {Moehler} S.,   {Neeser} M.~J.,  2013,
  \mn@doi [\aap] {10.1051/0004-6361/201322494}, \href
  {http://adsabs.harvard.edu/abs/2013A%26A...559A..96F} {559, A96}

\bibitem[\protect\citeauthoryear{{Fukugita}, {Ichikawa}, {Gunn}, {Doi},
  {Shimasaku}  \& {Schneider}}{{Fukugita} et~al.}{1996}]{Fukugita96}
{Fukugita} M.,  {Ichikawa} T.,  {Gunn} J.~E.,  {Doi} M.,  {Shimasaku} K.,
  {Schneider} D.~P.,  1996, \mn@doi [\aj] {10.1086/117915}, \href
  {http://adsabs.harvard.edu/abs/1996AJ....111.1748F} {111, 1748}

\bibitem[\protect\citeauthoryear{{Gehrels} et~al.,}{{Gehrels}
  et~al.}{2006}]{Gehrels06}
{Gehrels} N.,  et~al., 2006, \mn@doi [\nat] {10.1038/nature05376}, \href
  {https://ui.adsabs.harvard.edu/\#abs/2006Natur.444.1044G} {444, 1044}

\bibitem[\protect\citeauthoryear{{Gibson}, {Wynn}, {Gompertz}  \&
  {O'Brien}}{{Gibson} et~al.}{2017}]{Gibson17}
{Gibson} S.~L.,  {Wynn} G.~A.,  {Gompertz} B.~P.,   {O'Brien} P.~T.,  2017,
  \mn@doi [\mnras] {10.1093/mnras/stx1531}, \href
  {https://ui.adsabs.harvard.edu/#abs/2017MNRAS.470.4925G} {470, 4925}

\bibitem[\protect\citeauthoryear{{Gladders} \& {Yee}}{{Gladders} \&
  {Yee}}{2000}]{Gladders00}
{Gladders} M.~D.,  {Yee} H.~K.~C.,  2000, \mn@doi [\aj] {10.1086/301557}, \href
  {http://adsabs.harvard.edu/abs/2000AJ....120.2148G} {120, 2148}

\bibitem[\protect\citeauthoryear{{Glazebrook} \& {Bland-Hawthorn}}{{Glazebrook}
  \& {Bland-Hawthorn}}{2001}]{Glazebrook01}
{Glazebrook} K.,  {Bland-Hawthorn} J.,  2001, \mn@doi [\pasp] {10.1086/318625},
  \href {http://adsabs.harvard.edu/abs/2001PASP..113..197G} {113, 197}

\bibitem[\protect\citeauthoryear{{Goldstein} et~al.,}{{Goldstein}
  et~al.}{2017}]{Goldstein17}
{Goldstein} A.,  et~al., 2017, \mn@doi [\apjl] {10.3847/2041-8213/aa8f41},
  \href {http://adsabs.harvard.edu/abs/2017ApJ...848L..14G} {848, L14}

\bibitem[\protect\citeauthoryear{{Gompertz}, {O'Brien}, {Wynn}  \&
  {Rowlinson}}{{Gompertz} et~al.}{2013}]{Gompertz13}
{Gompertz} B.~P.,  {O'Brien} P.~T.,  {Wynn} G.~A.,   {Rowlinson} A.,  2013,
  \mn@doi [\mnras] {10.1093/mnras/stt293}, \href
  {http://adsabs.harvard.edu/abs/2013MNRAS.431.1745G} {431, 1745}

\bibitem[\protect\citeauthoryear{{Gompertz}, {O'Brien}  \& {Wynn}}{{Gompertz}
  et~al.}{2014}]{Gompertz14}
{Gompertz} B.~P.,  {O'Brien} P.~T.,   {Wynn} G.~A.,  2014, \mn@doi [\mnras]
  {10.1093/mnras/stt2165}, \href
  {http://adsabs.harvard.edu/abs/2014MNRAS.438..240G} {438, 240}

\bibitem[\protect\citeauthoryear{{Gompertz}, {van der Horst}, {O'Brien}, {Wynn}
   \& {Wiersema}}{{Gompertz} et~al.}{2015}]{Gompertz15}
{Gompertz} B.~P.,  {van der Horst} A.~J.,  {O'Brien} P.~T.,  {Wynn} G.~A.,
  {Wiersema} K.,  2015, \mn@doi [\mnras] {10.1093/mnras/stu2752}, \href
  {http://adsabs.harvard.edu/abs/2015MNRAS.448..629G} {448, 629}

\bibitem[\protect\citeauthoryear{{Gompertz} et~al.,}{{Gompertz}
  et~al.}{2018}]{Gompertz18}
{Gompertz} B.~P.,  et~al., 2018, \mn@doi [\apj] {10.3847/1538-4357/aac206},
  \href {http://adsabs.harvard.edu/abs/2018ApJ...860...62G} {860, 62}

\bibitem[\protect\citeauthoryear{{Granot} \& {Sari}}{{Granot} \&
  {Sari}}{2002}]{Granot02}
{Granot} J.,  {Sari} R.,  2002, \mn@doi [\apj] {10.1086/338966}, \href
  {https://ui.adsabs.harvard.edu/#abs/2002ApJ...568..820G} {568, 820}

\bibitem[\protect\citeauthoryear{{G{\"u}ver} \& {{\"O}zel}}{{G{\"u}ver} \&
  {{\"O}zel}}{2009}]{Guver09}
{G{\"u}ver} T.,  {{\"O}zel} F.,  2009, \mn@doi [\mnras]
  {10.1111/j.1365-2966.2009.15598.x}, \href
  {https://ui.adsabs.harvard.edu/abs/2009MNRAS.400.2050G} {400, 2050}

\bibitem[\protect\citeauthoryear{{Hamuy}, {Walker}, {Suntzeff}, {Gigoux},
  {Heathcote}  \& {Phillips}}{{Hamuy} et~al.}{1992}]{Hamuy92}
{Hamuy} M.,  {Walker} A.~R.,  {Suntzeff} N.~B.,  {Gigoux} P.,  {Heathcote}
  S.~R.,   {Phillips} M.~M.,  1992, \mn@doi [\pasp] {10.1086/133028}, \href
  {http://adsabs.harvard.edu/abs/1992PASP..104..533H} {104, 533}

\bibitem[\protect\citeauthoryear{{Hamuy}, {Suntzeff}, {Heathcote}, {Walker},
  {Gigoux}  \& {Phillips}}{{Hamuy} et~al.}{1994}]{Hamuy94}
{Hamuy} M.,  {Suntzeff} N.~B.,  {Heathcote} S.~R.,  {Walker} A.~R.,  {Gigoux}
  P.,   {Phillips} M.~M.,  1994, \mn@doi [\pasp] {10.1086/133417}, \href
  {http://adsabs.harvard.edu/abs/1994PASP..106..566H} {106, 566}

\bibitem[\protect\citeauthoryear{{Harrison} \& {Kobayashi}}{{Harrison} \&
  {Kobayashi}}{2013}]{Harrison13}
{Harrison} R.,  {Kobayashi} S.,  2013, \mn@doi [\apj]
  {10.1088/0004-637X/772/2/101}, \href
  {https://ui.adsabs.harvard.edu/\#abs/2013ApJ...772..101H} {772, 101}

\bibitem[\protect\citeauthoryear{{Hjorth} et~al.,}{{Hjorth}
  et~al.}{2003}]{Hjorth03}
{Hjorth} J.,  et~al., 2003, \mn@doi [\nat] {10.1038/nature01750}, \href
  {https://ui.adsabs.harvard.edu/#abs/2003Natur.423..847H} {423, 847}

\bibitem[\protect\citeauthoryear{{Hjorth} et~al.,}{{Hjorth}
  et~al.}{2005}]{Hjorth05b}
{Hjorth} J.,  et~al., 2005, \mn@doi [\nat] {10.1038/nature04174}, \href
  {http://adsabs.harvard.edu/abs/2005Natur.437..859H} {437, 859}

\bibitem[\protect\citeauthoryear{{Hook}, {J{\o}rgensen}, {Allington-Smith},
  {Davies}, {Metcalfe}, {Murowinski}  \& {Crampton}}{{Hook}
  et~al.}{2004}]{Hook04}
{Hook} I.~M.,  {J{\o}rgensen} I.,  {Allington-Smith} J.~R.,  {Davies} R.~L.,
  {Metcalfe} N.,  {Murowinski} R.~G.,   {Crampton} D.,  2004, \mn@doi [\pasp]
  {10.1086/383624}, \href {http://adsabs.harvard.edu/abs/2004PASP..116..425H}
  {116, 425}

\bibitem[\protect\citeauthoryear{{Jin} et~al.,}{{Jin} et~al.}{2016}]{Jin16}
{Jin} Z.-P.,  et~al., 2016, \mn@doi [Nature Communications]
  {10.1038/ncomms12898}, \href
  {http://adsabs.harvard.edu/abs/2016NatCo...712898J} {7, 12898}

\bibitem[\protect\citeauthoryear{{Jin}, {Covino}, {Liao}, {Li}, {D'Avanzo},
  {Fan}  \& {Wei}}{{Jin} et~al.}{2019}]{Jin19}
{Jin} Z.-P.,  {Covino} S.,  {Liao} N.-H.,  {Li} X.,  {D'Avanzo} P.,  {Fan}
  Y.-Z.,   {Wei} D.-M.,  2019, arXiv e-prints, \href
  {https://ui.adsabs.harvard.edu/\#abs/2019arXiv190106269J} {p.
  arXiv:1901.06269}

\bibitem[\protect\citeauthoryear{{Kasen}, {Fern{\'a}ndez}  \&
  {Metzger}}{{Kasen} et~al.}{2015}]{Kasen15}
{Kasen} D.,  {Fern{\'a}ndez} R.,   {Metzger} B.~D.,  2015, \mn@doi [\mnras]
  {10.1093/mnras/stv721}, \href
  {http://adsabs.harvard.edu/abs/2015MNRAS.450.1777K} {450, 1777}

\bibitem[\protect\citeauthoryear{{Kasliwal} et~al.,}{{Kasliwal}
  et~al.}{2017}]{Kasliwal17}
{Kasliwal} M.~M.,  et~al., 2017, \mn@doi [Science] {10.1126/science.aap9455},
  \href {http://adsabs.harvard.edu/abs/2017Sci...358.1559K} {358, 1559}

\bibitem[\protect\citeauthoryear{{Kennicutt}}{{Kennicutt}}{1998}]{Kennicutt98}
{Kennicutt} Jr. R.~C.,  1998, \mn@doi [\araa] {10.1146/annurev.astro.36.1.189},
  \href {http://adsabs.harvard.edu/abs/1998ARA%26A..36..189K} {36, 189}

\bibitem[\protect\citeauthoryear{{Kewley}, {Geller}  \& {Jansen}}{{Kewley}
  et~al.}{2004}]{Kewley04}
{Kewley} L.~J.,  {Geller} M.~J.,   {Jansen} R.~A.,  2004, \mn@doi [\aj]
  {10.1086/382723}, \href {http://adsabs.harvard.edu/abs/2004AJ....127.2002K}
  {127, 2002}

\bibitem[\protect\citeauthoryear{{Kilpatrick} et~al.,}{{Kilpatrick}
  et~al.}{2017}]{Kilpatrick17}
{Kilpatrick} C.~D.,  et~al., 2017, \mn@doi [Science] {10.1126/science.aaq0073},
  \href {https://ui.adsabs.harvard.edu/\#abs/2017Sci...358.1583K} {358, 1583}

\bibitem[\protect\citeauthoryear{{Kisaka}, {Ioka}  \& {Sakamoto}}{{Kisaka}
  et~al.}{2017}]{Kisaka17}
{Kisaka} S.,  {Ioka} K.,   {Sakamoto} T.,  2017, \mn@doi [\apj]
  {10.3847/1538-4357/aa8775}, \href
  {https://ui.adsabs.harvard.edu/abs/2017ApJ...846..142K} {846, 142}

\bibitem[\protect\citeauthoryear{{Kobayashi}}{{Kobayashi}}{2000}]{Kobayashi00b}
{Kobayashi} S.,  2000, \mn@doi [\apj] {10.1086/317869}, \href
  {https://ui.adsabs.harvard.edu/\#abs/2000ApJ...545..807K} {545, 807}

\bibitem[\protect\citeauthoryear{{Kobayashi} \& {Sari}}{{Kobayashi} \&
  {Sari}}{2000}]{Kobayashi00a}
{Kobayashi} S.,  {Sari} R.,  2000, \mn@doi [\apj] {10.1086/317021}, \href
  {https://ui.adsabs.harvard.edu/\#abs/2000ApJ...542..819K} {542, 819}

\bibitem[\protect\citeauthoryear{{Korobkin}, {Rosswog}, {Arcones}  \&
  {Winteler}}{{Korobkin} et~al.}{2012}]{Korobkin12}
{Korobkin} O.,  {Rosswog} S.,  {Arcones} A.,   {Winteler} C.,  2012, \mn@doi
  [\mnras] {10.1111/j.1365-2966.2012.21859.x}, \href
  {http://adsabs.harvard.edu/abs/2012MNRAS.426.1940K} {426, 1940}

\bibitem[\protect\citeauthoryear{{Kouveliotou}, {Meegan}, {Fishman}, {Bhat},
  {Briggs}, {Koshut}, {Paciesas}  \& {Pendleton}}{{Kouveliotou}
  et~al.}{1993}]{Kouveliotou93}
{Kouveliotou} C.,  {Meegan} C.~A.,  {Fishman} G.~J.,  {Bhat} N.~P.,  {Briggs}
  M.~S.,  {Koshut} T.~M.,  {Paciesas} W.~S.,   {Pendleton} G.~N.,  1993,
  \mn@doi [\apj] {10.1086/186969}, \href
  {https://ui.adsabs.harvard.edu/#abs/1993ApJ...413L.101K} {413, L101}

\bibitem[\protect\citeauthoryear{{Lagos}, {Bayet}, {Baugh}, {Lacey}, {Bell},
  {Fanidakis}  \& {Geach}}{{Lagos} et~al.}{2012}]{Lagos12}
{Lagos} C.~d.~P.,  {Bayet} E.,  {Baugh} C.~M.,  {Lacey} C.~G.,  {Bell} T.~A.,
  {Fanidakis} N.,   {Geach} J.~E.,  2012, \mn@doi [\mnras]
  {10.1111/j.1365-2966.2012.21905.x}, \href
  {http://adsabs.harvard.edu/abs/2012MNRAS.426.2142L} {426, 2142}

\bibitem[\protect\citeauthoryear{{Lamb} \& {Kobayashi}}{{Lamb} \&
  {Kobayashi}}{2016}]{Lamb16}
{Lamb} G.~P.,  {Kobayashi} S.,  2016, \mn@doi [\apj]
  {10.3847/0004-637X/829/2/112}, \href
  {https://ui.adsabs.harvard.edu/\#abs/2016ApJ...829..112L} {829, 112}

\bibitem[\protect\citeauthoryear{{Lamb} et~al.,}{{Lamb} et~al.}{2019}]{Lamb19}
{Lamb} G.~P.,  et~al., 2019, arXiv e-prints, \href
  {https://ui.adsabs.harvard.edu/abs/2019arXiv190502159L} {p. arXiv:1905.02159}

\bibitem[\protect\citeauthoryear{{Lattimer} \& {Schramm}}{{Lattimer} \&
  {Schramm}}{1974}]{Lattimer74}
{Lattimer} J.~M.,  {Schramm} D.~N.,  1974, \mn@doi [\apjl] {10.1086/181612},
  \href {http://adsabs.harvard.edu/abs/1974ApJ...192L.145L} {192, L145}

\bibitem[\protect\citeauthoryear{{Leibler} \& {Berger}}{{Leibler} \&
  {Berger}}{2010}]{Leibler10}
{Leibler} C.~N.,  {Berger} E.,  2010, \mn@doi [\apj]
  {10.1088/0004-637X/725/1/1202}, \href
  {http://adsabs.harvard.edu/abs/2010ApJ...725.1202L} {725, 1202}

\bibitem[\protect\citeauthoryear{{Li} \& {Paczy{\'n}ski}}{{Li} \&
  {Paczy{\'n}ski}}{1998}]{Li98}
{Li} L.-X.,  {Paczy{\'n}ski} B.,  1998, \mn@doi [\apj] {10.1086/311680}, \href
  {https://ui.adsabs.harvard.edu/\#abs/1998ApJ...507L..59L} {507, L59}

\bibitem[\protect\citeauthoryear{{Li}, {Li}  \& {Wei}}{{Li}
  et~al.}{2008a}]{Li08a}
{Li} Y.,  {Li} A.,   {Wei} D.~M.,  2008a, \mn@doi [\apj] {10.1086/528734},
  \href {https://ui.adsabs.harvard.edu/abs/2008ApJ...678.1136L} {678, 1136}

\bibitem[\protect\citeauthoryear{{Li}, {Liang}, {Kann}, {Wei}, {Klose}  \&
  {Wang}}{{Li} et~al.}{2008b}]{Li08b}
{Li} A.,  {Liang} S.~L.,  {Kann} D.~A.,  {Wei} D.~M.,  {Klose} S.,   {Wang}
  Y.~J.,  2008b, \mn@doi [\apj] {10.1086/591049}, \href
  {https://ui.adsabs.harvard.edu/abs/2008ApJ...685.1046L} {685, 1046}

\bibitem[\protect\citeauthoryear{{Lien} et~al.,}{{Lien} et~al.}{2016}]{Lien16}
{Lien} A.,  et~al., 2016, \mn@doi [\apj] {10.3847/0004-637X/829/1/7}, \href
  {http://adsabs.harvard.edu/abs/2016ApJ...829....7L} {829, 7}

\bibitem[\protect\citeauthoryear{{Lyman} et~al.,}{{Lyman}
  et~al.}{2018}]{Lyman18b}
{Lyman} J.~D.,  et~al., 2018, \mn@doi [\mnras] {10.1093/mnras/stx2414}, \href
  {http://adsabs.harvard.edu/abs/2018MNRAS.473.1359L} {473, 1359}

\bibitem[\protect\citeauthoryear{{Malesani} et~al.,}{{Malesani}
  et~al.}{2007}]{Malesani07}
{Malesani} D.,  et~al., 2007, \mn@doi [\aap] {10.1051/0004-6361:20077868},
  \href {https://ui.adsabs.harvard.edu/#abs/2007A&A...473...77M} {473, 77}

\bibitem[\protect\citeauthoryear{{Mannucci}, {Salvaterra}  \&
  {Campisi}}{{Mannucci} et~al.}{2011}]{Mannucci11}
{Mannucci} F.,  {Salvaterra} R.,   {Campisi} M.~A.,  2011, \mn@doi [\mnras]
  {10.1111/j.1365-2966.2011.18459.x}, \href
  {http://adsabs.harvard.edu/abs/2011MNRAS.414.1263M} {414, 1263}

\bibitem[\protect\citeauthoryear{{Margutti} et~al.,}{{Margutti}
  et~al.}{2011}]{Margutti11}
{Margutti} R.,  et~al., 2011, \mn@doi [\mnras]
  {10.1111/j.1365-2966.2011.19397.x}, \href
  {http://adsabs.harvard.edu/abs/2011MNRAS.417.2144M} {417, 2144}

\bibitem[\protect\citeauthoryear{{McBreen} et~al.,}{{McBreen}
  et~al.}{2008}]{McBreen08}
{McBreen} S.,  et~al., 2008, \mn@doi [\apj] {10.1086/588189}, \href
  {https://ui.adsabs.harvard.edu/\#abs/2008ApJ...677L..85M} {677, L85}

\bibitem[\protect\citeauthoryear{{McMillan}}{{McMillan}}{2017}]{McMillan17}
{McMillan} P.~J.,  2017, \mn@doi [\mnras] {10.1093/mnras/stw2759}, \href
  {http://adsabs.harvard.edu/abs/2017MNRAS.465...76M} {465, 76}

\bibitem[\protect\citeauthoryear{Metzger}{Metzger}{2017}]{Metzger17}
Metzger B.~D.,  2017, \mn@doi [Living Reviews in Relativity]
  {10.1007/s41114-017-0006-z}, 20

\bibitem[\protect\citeauthoryear{{Metzger} \& {Fern{\'a}ndez}}{{Metzger} \&
  {Fern{\'a}ndez}}{2014}]{Metzger14b}
{Metzger} B.~D.,  {Fern{\'a}ndez} R.,  2014, \mn@doi [\mnras]
  {10.1093/mnras/stu802}, \href
  {http://adsabs.harvard.edu/abs/2014MNRAS.441.3444M} {441, 3444}

\bibitem[\protect\citeauthoryear{{Metzger} \& {Piro}}{{Metzger} \&
  {Piro}}{2014}]{Metzger14a}
{Metzger} B.~D.,  {Piro} A.~L.,  2014, \mn@doi [\mnras] {10.1093/mnras/stu247},
  \href {http://adsabs.harvard.edu/abs/2014MNRAS.439.3916M} {439, 3916}

\bibitem[\protect\citeauthoryear{{Metzger}, {Bauswein}, {Goriely}  \&
  {Kasen}}{{Metzger} et~al.}{2015}]{Metzger15}
{Metzger} B.~D.,  {Bauswein} A.,  {Goriely} S.,   {Kasen} D.,  2015, \mn@doi
  [\mnras] {10.1093/mnras/stu2225}, \href
  {http://adsabs.harvard.edu/abs/2015MNRAS.446.1115M} {446, 1115}

\bibitem[\protect\citeauthoryear{{M{\"o}ller}, {Nix}, {Myers}  \&
  {Swiatecki}}{{M{\"o}ller} et~al.}{1995}]{Moller95}
{M{\"o}ller} P.,  {Nix} J.~R.,  {Myers} W.~D.,   {Swiatecki} W.~J.,  1995,
  \mn@doi [Atomic Data and Nuclear Data Tables] {10.1006/adnd.1995.1002}, \href
  {https://ui.adsabs.harvard.edu/#abs/1995ADNDT..59..185M} {59, 185}

\bibitem[\protect\citeauthoryear{{Nicuesa Guelbenzu} et~al.,}{{Nicuesa
  Guelbenzu} et~al.}{2012}]{NicuesaGuelbenzu12}
{Nicuesa Guelbenzu} A.,  et~al., 2012, \mn@doi [\aap]
  {10.1051/0004-6361/201219551}, \href
  {https://ui.adsabs.harvard.edu/#abs/2012A&A...548A.101N} {548, A101}

\bibitem[\protect\citeauthoryear{Nicuesa~Guelbenzu et~al.,}{Nicuesa~Guelbenzu
  et~al.}{2014}]{NicuesaGuelbenzu14}
Nicuesa~Guelbenzu A.,  et~al., 2014, \mn@doi [\apj]
  {10.1088/0004-637X/789/1/45}, \href
  {http://adsabs.harvard.edu/abs/2014ApJ...789...45N} {789, 45}

\bibitem[\protect\citeauthoryear{{Norris}, {Gehrels}  \& {Scargle}}{{Norris}
  et~al.}{2010}]{Norris10}
{Norris} J.~P.,  {Gehrels} N.,   {Scargle} J.~D.,  2010, \mn@doi [\apj]
  {10.1088/0004-637X/717/1/411}, \href
  {https://ui.adsabs.harvard.edu/#abs/2010ApJ...717..411N} {717, 411}

\bibitem[\protect\citeauthoryear{{Osterbrock} \& {Ferland}}{{Osterbrock} \&
  {Ferland}}{2006}]{Osterbrock06}
{Osterbrock} D.~E.,  {Ferland} G.~J.,  2006, {Astrophysics of gaseous nebulae
  and active galactic nuclei}

\bibitem[\protect\citeauthoryear{{Pandey} et~al.,}{{Pandey}
  et~al.}{2019}]{Pandey19}
{Pandey} S.~B.,  et~al., 2019, \mn@doi [\mnras] {10.1093/mnras/stz530}, \href
  {https://ui.adsabs.harvard.edu/abs/2019MNRAS.485.5294P} {485, 5294}

\bibitem[\protect\citeauthoryear{{Pei}}{{Pei}}{1992}]{XSPECdust}
{Pei} Y.~C.,  1992, \mn@doi [\apj] {10.1086/171637}, \href
  {http://adsabs.harvard.edu/abs/1992ApJ...395..130P} {395, 130}

\bibitem[\protect\citeauthoryear{{Perley} et~al.,}{{Perley}
  et~al.}{2009}]{Perley09}
{Perley} D.~A.,  et~al., 2009, \mn@doi [\apj] {10.1088/0004-637X/696/2/1871},
  \href {http://adsabs.harvard.edu/abs/2009ApJ...696.1871P} {696, 1871}

\bibitem[\protect\citeauthoryear{{Perna}, {Armitage}  \& {Zhang}}{{Perna}
  et~al.}{2006}]{Perna06}
{Perna} R.,  {Armitage} P.~J.,   {Zhang} B.,  2006, \mn@doi [\apj]
  {10.1086/499775}, \href
  {https://ui.adsabs.harvard.edu/#abs/2006ApJ...636L..29P} {636, L29}

\bibitem[\protect\citeauthoryear{{Pettini} \& {Pagel}}{{Pettini} \&
  {Pagel}}{2004}]{Pettini04}
{Pettini} M.,  {Pagel} B.~E.~J.,  2004, \mn@doi [\mnras]
  {10.1111/j.1365-2966.2004.07591.x}, \href
  {http://adsabs.harvard.edu/abs/2004MNRAS.348L..59P} {348, L59}

\bibitem[\protect\citeauthoryear{{Piran}}{{Piran}}{2004}]{Piran04}
{Piran} T.,  2004, \mn@doi [Reviews of Modern Physics]
  {10.1103/RevModPhys.76.1143}, \href
  {https://ui.adsabs.harvard.edu/#abs/2004RvMP...76.1143P} {76, 1143}

\bibitem[\protect\citeauthoryear{{Rieke} \& {Lebofsky}}{{Rieke} \&
  {Lebofsky}}{1985}]{Rieke85}
{Rieke} G.~H.,  {Lebofsky} M.~J.,  1985, \mn@doi [\apj] {10.1086/162827}, \href
  {http://adsabs.harvard.edu/abs/1985ApJ...288..618R} {288, 618}

\bibitem[\protect\citeauthoryear{{Rossi} et~al.,}{{Rossi}
  et~al.}{2019}]{Rossi19}
{Rossi} A.,  et~al., 2019, arXiv e-prints, \href
  {https://ui.adsabs.harvard.edu/abs/2019arXiv190105792R} {p. arXiv:1901.05792}

\bibitem[\protect\citeauthoryear{{Rosswog}}{{Rosswog}}{2005}]{Rosswog05}
{Rosswog} S.,  2005, \mn@doi [\apj] {10.1086/497062}, \href
  {http://cdsads.u-strasbg.fr/abs/2005ApJ...634.1202R} {634, 1202}

\bibitem[\protect\citeauthoryear{{Rosswog}}{{Rosswog}}{2007}]{Rosswog07}
{Rosswog} S.,  2007, \mn@doi [\mnras] {10.1111/j.1745-3933.2007.00284.x}, \href
  {https://ui.adsabs.harvard.edu/#abs/2007MNRAS.376L..48R} {376, L48}

\bibitem[\protect\citeauthoryear{{Rosswog}}{{Rosswog}}{2015}]{Rosswog15}
{Rosswog} S.,  2015, \mn@doi [International Journal of Modern Physics D]
  {10.1142/S0218271815300128}, \href
  {https://ui.adsabs.harvard.edu/#abs/2015IJMPD..2430012R} {24, 1530012}

\bibitem[\protect\citeauthoryear{{Rosswog}, {Feindt}, {Korobkin}, {Wu},
  {Sollerman}, {Goobar}  \& {Martinez-Pinedo}}{{Rosswog}
  et~al.}{2017}]{Rosswog17}
{Rosswog} S.,  {Feindt} U.,  {Korobkin} O.,  {Wu} M.-R.,  {Sollerman} J.,
  {Goobar} A.,   {Martinez-Pinedo} G.,  2017, \mn@doi [Classical and Quantum
  Gravity] {10.1088/1361-6382/aa68a9}, \href
  {http://adsabs.harvard.edu/abs/2017CQGra..34j4001R} {34, 104001}

\bibitem[\protect\citeauthoryear{{Rowlinson} et~al.,}{{Rowlinson}
  et~al.}{2010}]{Rowlinson10}
{Rowlinson} A.,  et~al., 2010, \mn@doi [\mnras]
  {10.1111/j.1365-2966.2010.17115.x}, \href
  {https://ui.adsabs.harvard.edu/\#abs/2010MNRAS.408..383R} {408, 383}

\bibitem[\protect\citeauthoryear{{Sakamoto} et~al.,}{{Sakamoto}
  et~al.}{2007a}]{Sakamoto07a}
{Sakamoto} T.,  et~al., 2007a, GRB Coordinates Network, \href
  {http://adsabs.harvard.edu/abs/2007GCN..7147....1S} {7147}

\bibitem[\protect\citeauthoryear{{Sakamoto}, {Norris}, {Ukwatta}, {Barthelmy},
  {Gehrels}  \& {Stamatikos}}{{Sakamoto} et~al.}{2007b}]{Sakamoto07b}
{Sakamoto} T.,  {Norris} J.,  {Ukwatta} T.,  {Barthelmy} S.~D.,  {Gehrels} N.,
   {Stamatikos} M.,  2007b, GRB Coordinates Network, \href
  {http://adsabs.harvard.edu/abs/2007GCN..7156....1S} {7156}

\bibitem[\protect\citeauthoryear{{Sari} \& {Piran}}{{Sari} \&
  {Piran}}{1999}]{Sari99b}
{Sari} R.,  {Piran} T.,  1999, \mn@doi [\apj] {10.1086/307508}, \href
  {https://ui.adsabs.harvard.edu/#abs/1999ApJ...520..641S} {520, 641}

\bibitem[\protect\citeauthoryear{{Sari}, {Piran}  \& {Narayan}}{{Sari}
  et~al.}{1998}]{Sari98}
{Sari} R.,  {Piran} T.,   {Narayan} R.,  1998, \mn@doi [\apj] {10.1086/311269},
  \href {https://ui.adsabs.harvard.edu/#abs/1998ApJ...497L..17S} {497, L17}

\bibitem[\protect\citeauthoryear{{Sari}, {Piran}  \& {Halpern}}{{Sari}
  et~al.}{1999}]{Sari99a}
{Sari} R.,  {Piran} T.,   {Halpern} J.~P.,  1999, \mn@doi [\apj]
  {10.1086/312109}, \href
  {https://ui.adsabs.harvard.edu/#abs/1999ApJ...519L..17S} {519, L17}

\bibitem[\protect\citeauthoryear{{Savaglio}, {Glazebrook}  \& {Le
  Borgne}}{{Savaglio} et~al.}{2009}]{Savaglio09}
{Savaglio} S.,  {Glazebrook} K.,   {Le Borgne} D.,  2009, \mn@doi [\apj]
  {10.1088/0004-637X/691/1/182}, \href
  {https://ui.adsabs.harvard.edu/abs/2009ApJ...691..182S} {691, 182}

\bibitem[\protect\citeauthoryear{{Savchenko} et~al.,}{{Savchenko}
  et~al.}{2017}]{Savchenko17}
{Savchenko} V.,  et~al., 2017, \mn@doi [\apjl] {10.3847/2041-8213/aa8f94},
  \href {http://adsabs.harvard.edu/abs/2017ApJ...848L..15S} {848, L15}

\bibitem[\protect\citeauthoryear{{Schlegel}, {Finkbeiner}  \&
  {Davis}}{{Schlegel} et~al.}{1998}]{Schlegel98}
{Schlegel} D.~J.,  {Finkbeiner} D.~P.,   {Davis} M.,  1998, \mn@doi [\apj]
  {10.1086/305772}, \href {http://adsabs.harvard.edu/abs/1998ApJ...500..525S}
  {500, 525}

\bibitem[\protect\citeauthoryear{Shaw}{Shaw}{2016}]{Shaw16}
Shaw R.~A.,  2016, GMOS Data Reduction Cookbook, \url
  {http://ast.noao.edu/sites/default/files/GMOS_Cookbook/}

\bibitem[\protect\citeauthoryear{{Smartt} et~al.,}{{Smartt}
  et~al.}{2017}]{Smartt17}
{Smartt} S.~J.,  et~al., 2017, \mn@doi [\nat] {10.1038/nature24303}, \href
  {https://ui.adsabs.harvard.edu/\#abs/2017Natur.551...75S} {551, 75}

\bibitem[\protect\citeauthoryear{{Springel} et~al.,}{{Springel}
  et~al.}{2005}]{Springel05}
{Springel} V.,  et~al., 2005, \mn@doi [\nat] {10.1038/nature03597}, \href
  {https://ui.adsabs.harvard.edu/#abs/2005Natur.435..629S} {435, 629}

\bibitem[\protect\citeauthoryear{{Stott}, {Pimbblet}, {Edge}, {Smith}  \&
  {Wardlow}}{{Stott} et~al.}{2009}]{Stott09}
{Stott} J.~P.,  {Pimbblet} K.~A.,  {Edge} A.~C.,  {Smith} G.~P.,   {Wardlow}
  J.~L.,  2009, \mn@doi [\mnras] {10.1111/j.1365-2966.2009.14477.x}, \href
  {http://adsabs.harvard.edu/abs/2009MNRAS.394.2098S} {394, 2098}

\bibitem[\protect\citeauthoryear{{Svensson}, {Levan}, {Tanvir}, {Fruchter}  \&
  {Strolger}}{{Svensson} et~al.}{2010}]{Svensson10}
{Svensson} K.~M.,  {Levan} A.~J.,  {Tanvir} N.~R.,  {Fruchter} A.~S.,
  {Strolger} L.~G.,  2010, \mn@doi [\mnras] {10.1111/j.1365-2966.2010.16442.x},
  \href {https://ui.adsabs.harvard.edu/#abs/2010MNRAS.405...57S} {405, 57}

\bibitem[\protect\citeauthoryear{{Tanga}, {Kr{\"u}hler}, {Schady}, {Klose},
  {Graham}, {Greiner}, {Kann}  \& {Nardini}}{{Tanga} et~al.}{2018}]{Tanga18}
{Tanga} M.,  {Kr{\"u}hler} T.,  {Schady} P.,  {Klose} S.,  {Graham} J.~F.,
  {Greiner} J.,  {Kann} D.~A.,   {Nardini} M.,  2018, \mn@doi [\aap]
  {10.1051/0004-6361/201731799}, \href
  {https://ui.adsabs.harvard.edu/\#abs/2018A&A...615A.136T} {615, A136}

\bibitem[\protect\citeauthoryear{{Tanvir}, {Levan}, {Fruchter}, {Hjorth},
  {Hounsell}, {Wiersema}  \& {Tunnicliffe}}{{Tanvir} et~al.}{2013}]{Tanvir13}
{Tanvir} N.~R.,  {Levan} A.~J.,  {Fruchter} A.~S.,  {Hjorth} J.,  {Hounsell}
  R.~A.,  {Wiersema} K.,   {Tunnicliffe} R.~L.,  2013, \mn@doi [\nat]
  {10.1038/nature12505}, \href
  {http://adsabs.harvard.edu/abs/2013Natur.500..547T} {500, 547}

\bibitem[\protect\citeauthoryear{{Tanvir} et~al.,}{{Tanvir}
  et~al.}{2017}]{Tanvir17}
{Tanvir} N.~R.,  et~al., 2017, \mn@doi [\apjl] {10.3847/2041-8213/aa90b6},
  \href {http://adsabs.harvard.edu/abs/2017ApJ...848L..27T} {848, L27}

\bibitem[\protect\citeauthoryear{{Troja} et~al.,}{{Troja}
  et~al.}{2018}]{Troja18b}
{Troja} E.,  et~al., 2018, \mn@doi [Nature Communications]
  {10.1038/s41467-018-06558-7}, \href
  {https://ui.adsabs.harvard.edu/abs/2018NatCo...9.4089T} {9, 4089}

\bibitem[\protect\citeauthoryear{{Troja} et~al.,}{{Troja}
  et~al.}{2019}]{Troja19}
{Troja} E.,  et~al., 2019, arXiv e-prints, \href
  {https://ui.adsabs.harvard.edu/abs/2019arXiv190501290T} {p. arXiv:1905.01290}

\bibitem[\protect\citeauthoryear{{Walter}, {Brinks}, {de Blok}, {Bigiel},
  {Kennicutt}, {Thornley}  \& {Leroy}}{{Walter} et~al.}{2008}]{Walter08}
{Walter} F.,  {Brinks} E.,  {de Blok} W.~J.~G.,  {Bigiel} F.,  {Kennicutt}
  Robert~C. J.,  {Thornley} M.~D.,   {Leroy} A.,  2008, \mn@doi [\aj]
  {10.1088/0004-6256/136/6/2563}, \href
  {https://ui.adsabs.harvard.edu/\#abs/2008AJ....136.2563W} {136, 2563}

\bibitem[\protect\citeauthoryear{{Willingale}, {Starling}, {Beardmore},
  {Tanvir}  \& {O'Brien}}{{Willingale} et~al.}{2013}]{XSPECgalcolumn}
{Willingale} R.,  {Starling} R.~L.~C.,  {Beardmore} A.~P.,  {Tanvir} N.~R.,
  {O'Brien} P.~T.,  2013, \mn@doi [\mnras] {10.1093/mnras/stt175}, \href
  {https://ui.adsabs.harvard.edu/#abs/2013MNRAS.431..394W} {431, 394}

\bibitem[\protect\citeauthoryear{{Wilms}, {Allen}  \& {McCray}}{{Wilms}
  et~al.}{2000}]{XSPECtbabs}
{Wilms} J.,  {Allen} A.,   {McCray} R.,  2000, \mn@doi [\apj] {10.1086/317016},
  \href {http://adsabs.harvard.edu/abs/2000ApJ...542..914W} {542, 914}

\bibitem[\protect\citeauthoryear{{Wolf} et~al.,}{{Wolf} et~al.}{2018}]{Wolf18}
{Wolf} C.,  et~al., 2018, \mn@doi [\pasa] {10.1017/pasa.2018.5}, \href
  {http://adsabs.harvard.edu/abs/2018PASA...35...10W} {35, e010}

\bibitem[\protect\citeauthoryear{{Xu} et~al.,}{{Xu} et~al.}{2009}]{Xu09}
{Xu} D.,  et~al., 2009, \mn@doi [\apj] {10.1088/0004-637X/696/1/971}, \href
  {https://ui.adsabs.harvard.edu/#abs/2009ApJ...696..971X} {696, 971}

\bibitem[\protect\citeauthoryear{{Yang} et~al.,}{{Yang} et~al.}{2015}]{Yang15}
{Yang} B.,  et~al., 2015, \mn@doi [Nature Communications] {10.1038/ncomms8323},
  \href {http://adsabs.harvard.edu/abs/2015NatCo...6E7323Y} {6, 7323}

\bibitem[\protect\citeauthoryear{{Yoshida}, {Yonetoku}, {Arimoto}, {Sawano}  \&
  {Kagawa}}{{Yoshida} et~al.}{2019}]{Yoshida19}
{Yoshida} K.,  {Yonetoku} D.,  {Arimoto} M.,  {Sawano} T.,   {Kagawa} Y.,
  2019, \mn@doi [\pasj] {10.1093/pasj/psz030}, \href
  {https://ui.adsabs.harvard.edu/abs/2019PASJ..tmp...47Y} {p.~47}

\bibitem[\protect\citeauthoryear{{de Ugarte Postigo} et~al.,}{{de Ugarte
  Postigo} et~al.}{2014}]{deUgartePostigo14}
{de Ugarte Postigo} A.,  et~al., 2014, \mn@doi [\aap]
  {10.1051/0004-6361/201322985}, \href
  {https://ui.adsabs.harvard.edu/abs/2014A&A...563A..62D} {563, A62}

\makeatother
\end{thebibliography}



\appendix

\section{Spectral energy distribution (SED) fits}
\label{app:SED_detail}

\begin{table*}
\centering
\caption{Our \textsc{xspec} fit to the X-ray spectrum using the absorbed power law model \texttt{tbabs*ztbabs(powerlaw)}. This model represents a power law spectrum, where flux density $A(E)=KE^{-\Gamma}$ for some energy $E$, $\Gamma$ is the photon index, with $\Gamma=\beta+1$, and $K$ is the flux density at 1 keV. \texttt{Tbabs} and \texttt{ztbabs} apply interstellar medium (ISM) absorption corrections to the X-ray, with \texttt{tbabs} correcting for the Milky Way and \texttt{ztbabs} correcting for the host and its redshift \citep{XSPECtbabs}. All errors are given to 90\% confidence limits.}
\label{tab:Xray_spec}
\begin{tabular}{@{}cccc@{}}
\hline
Model Component & Parameter & Value & Free Parameter? \\
\hline
\texttt{TBabs} & $N_H$ & $1.3 \times 10^{20}$ cm$^{-2}$ & N \\
\texttt{zTBabs} & $N_H$ & $\leq 2.94 \times 10^{21}$ cm$^{-2}$ & Y \\
\texttt{zTBabs} & $z$ & 0.381 & N \\
\texttt{powerlaw} & $\Gamma$ & $1.58^{+0.87}_{-0.57}$ & Y \\
\texttt{powerlaw} & $K$ & $4.81^{+4.43}_{-1.75}\times 10^{-6}$ photons keV$^{-1}$ cm$^{-2}$ s$^{-1}$ & Y \\
\hline
\end{tabular}
\end{table*}

\begin{table*}
\centering
\caption{Our \textsc{xspec} fit to the first epoch Gemini optical data using the absorbed power law model \texttt{redden*zdust(powerlaw)}, where \texttt{redden} introduces a correction for IR/optical/UV extinction in the Milky Way \citep{Cardelli89} while \texttt{zdust} corrects for dust extinction in the host, again including its redshift \citep{XSPECdust}. Note that the \texttt{method} selected uses the extinction curves derived from the Milky Way. All errors are given to 90\% confidence limits.}
\label{tab:Optical_spec}
\begin{tabular}{@{}cccc@{}}
\hline
Model Component & Parameter & Value & Free Parameter? \\
\hline
\texttt{redden} & $E(B-V)$ & 0.013 & N \\
\texttt{zdust} & \texttt{method} & 1 & N \\
\texttt{zdust} & $E(B-V)$ & 0.00 & N \\
\texttt{zdust} & $R_v$ & 3.08 & N \\
\texttt{zdust} & $z$ & 0.381 & N \\
\texttt{powerlaw} & $\Gamma$ & $\leq 4.49$ & Y \\
\texttt{powerlaw} & $K$ & $4.93^{+1.56\times10^4}_{-4.93}\times 10^{-13}$ photons keV$^{-1}$ cm$^{-2}$ s$^{-1}$ & Y \\
\hline
\end{tabular}
\end{table*}

\begin{table*}
\centering
\caption{Our \textsc{xspec} fit to the combined X-ray and optical SED using the absorbed power law model \texttt{redden*tbabs(zdust*ztbabs(powerlaw))}. As the normalisation of the power law was allowed to vary between the X-ray and optical data, we list both below. All errors are given to 90\% confidence limits.}
\label{tab:Combined_spec}
\begin{tabular}{@{}cccc@{}}
\hline
Model Component & Parameter & Value & Free Parameter? \\
\hline
\texttt{redden} & $E(B-V)$ & 0.013 & N \\
\texttt{TBabs} & $N_H$ & $1.3 \times 10^{20}$ cm$^{-2}$ & N \\
\texttt{zdust} & \texttt{method} & 1 & N \\
\texttt{zdust} & $E(B-V)$ & $\leq$0.85 & Y \\
\texttt{zdust} & $R_v$ & 3.08 & N \\
\texttt{zdust} & $z$ & 0.381 & N \\
\texttt{zTBabs} & $N_H$ & $\leq 8.22 \times 10^{21}$ cm$^{-2}$ & Y \\
\texttt{zTBabs} & $z$ & 0.381 & N \\
\texttt{powerlaw} & $\Gamma$ & $1.58^{+0.56}_{-0.57}$ & Y \\
\texttt{powerlaw} & $K_{\text{X-ray}}$ & $4.81^{+2.13}_{-1.75}\times 10^{-6}$ keV$^{-1}$ cm$^{-2}$ s$^{-1}$ & Y \\
\texttt{powerlaw} & $K_{\text{Optical}}$ & $\leq6.69\times 10^{-4}$ photons keV$^{-1}$ cm$^{-2}$ s$^{-1}$ & Y \\
\hline
\end{tabular}
\end{table*}

\begin{table*}
\centering
\caption{Our Markov Chain Monte Carlo (MCMC) power law fit to the afterglow defined as $A(E)=KE^{-\Gamma}$ for some energy $E$ as above. $E(B-V)$ is derived from the $N_H$ using the relation identified by \citet{Guver09}. All errors below are 1 $\sigma$.}
\label{tab:MCMC_tied}
\begin{tabular}{@{}cc@{}}
\hline
Parameter & Value \\
\hline
$\Gamma$ & $2.19^{+0.06}_{-0.04}$\\
$K$ & $4.12^{0.94}_{-1.05} \times 10^{-6}$ photons keV$^{-1}$ cm$^{-2}$ s$^{-1}$\\
$N_H$ & $2.97^{+0.01}_{-0.64} \times 10^{21}$ cm$^{-2}$\\
$E(B-V)$ & $0.43^{+0.01}_{-0.09}$ mag\\
\hline
\end{tabular}
\end{table*}

\begin{table*}
\centering
\caption{Our MCMC power law fit to the afterglow where $E(B-V)$ is not linked explicitly to the $N_H$ and all parameters are free to vary. All errors below are 1 $\sigma$.}
\label{tab:MCMC_untied}
\begin{tabular}{@{}cc@{}}
\hline
Parameter & Value \\
\hline
$\Gamma$ & $2.15^{+0.01}_{-0.06}$\\
$K$ & $5.80^{+1.70}_{-0.04} \times 10^{-6}$ photons keV$^{-1}$ cm$^{-2}$ s$^{-1}$\\
$N_H$ & $2.77^{+5.51}_{-2.77} \times 10^{20}$ cm$^{-2}$\\
$E(B-V)$ & $0.45^{+0.01}_{-0.07}$ mag\\
\hline
\end{tabular}
\end{table*}




\bsp	
\label{lastpage}
\end{document}